\documentclass[11pt]{article}
\usepackage[utf8]{inputenc}
\usepackage[a4paper, margin=2.5cm]{geometry}
\usepackage{graphicx}
\usepackage{xcolor}
 \usepackage{float} 
\usepackage{latexsym}
\usepackage{comment}
\usepackage{multicol}
\usepackage{dsfont}
\usepackage{braket}
\usepackage{newunicodechar}
\newunicodechar{⁻}{$^{-}$}
\usepackage{amssymb,amsthm, amsmath}
\usepackage{quantikz}
\usepackage{hyperref}
\usepackage{subcaption}
\usepackage{caption}
\usepackage{mathtools} 
\usepackage{dsfont}
\usepackage{booktabs}
\usepackage{array}
\usepackage[backend=bibtex,sorting=none,giveninits=true,maxnames=99,maxcitenames=99]{biblatex}
\usepackage{soul}
\usepackage{authblk}
\usepackage{enumitem}
\usepackage{makecell}
\usepackage[normalem]{ulem}

\title{
Coupled integrated photonic quantum memristors\\ using a single photon source made of a colour center}

\author[1,*]{Alessio Baldazzi}
\author[2,*]{Roy Philip George Konnoth Ancel}
\author[1]{Sebastiano Guaraldo}
\author[1]{Ivan Fattori}
\author[1]{Xuan Chen}
\author[2]{Ziad Abi Akar}
\author[2]{Regis Deturche}
\author[1]{Stefano Azzini}
\author[2]{Christophe Couteau}
\author[1]{Lorenzo Pavesi}

\affil[1]{Department of Physics, University of Trento, Via Sommarive 14, Trento, 38123, Italy}
\affil[2]{Laboratory Light, nanomaterials and nanotechnologies - L2n UMR 7076, University of Technology of Troyes-UTT and CNRS, 12 Rue Marie Curie, 10300 Troyes, France}
\affil[*]{These authors contributed equally to this work.}

\date{}

\begin{document}

\maketitle

\begin{abstract}
Photonic quantum memristors provide a measurement-induced route to nonlinear and history-dependent quantum dynamics. Experimental demonstrations have so far focused on isolated devices or simple cascaded devices configurations. Here, we experimentally realize and characterize a network of two coupled photonic quantum memristors with crossed feedback, implemented on a silicon nitride photonic integrated circuit and fed by a room-temperature single-photon source based on a silicon-vacancy color center SiV$^-$ in a nanodiamond. Each memristor consists of an integrated Mach-Zehnder interferometer whose transfer function is adaptively updated by photon detection events on another memristor, thus generating novel non-Markovian input-output dynamics with an enhanced memristive behaviour compared to single devices. In particular, we report inter-memristor input-output hysteresis curves exhibiting larger form factors and displaying self-intersecting loops, respectively revealing marked bistability and self-intersecting hysteresis geometry. Furthermore, numerical simulations show how these features emerge from the interplay between memory depth and relative input phase, for both intra- and inter-memristor input-output relations. We experimentally test the performance of our system in the NARMA task. Our results establish coupled integrated photonic quantum memristors as scalable nonlinear building blocks and highlight their potential for implementing compact quantum neuromorphic and reservoir computing architectures.


\end{abstract}


\section{Introduction}

Machine learning has become one of the most powerful tools in industry, business, and daily life, enabling efficient data processing and decision-making~\cite{lecun2015deep,ismail2019deep,pierson2017deep}.
Building on these successes, it is also increasingly adopted as an innovative tool in scientific research~\cite{huang2022quantum,PhysRevLett.120.240501,carleo2019machine}. A key ingredient in machine learning is the presence of nonlinear transformations together with adaptive, state-dependent responses that provide memory, enabling learning and complex information processing. As efforts intensify to transfer these capabilities to the quantum domain, quantum machine learning has emerged as a promising direction with advantages for a broad class of problems, in particular for tasks that involve quantum data or quantum-generated correlations~\cite{biamonte2017quantum,du2025efficient,mo2025parameterized,jerbi2024shadows,saggio2021experimental}. Yet, conventional quantum evolution is naturally unitary, and therefore inherently linear. As a result, realizing effective nonlinear dynamics typically relies on measurement-induced processes, open quantum systems, or feedback-based approaches~\cite{knill2001scheme,scheel2003measurement,zhou2006nonlinear}. Especially in quantum photonic systems, nonlinear interactions at the level of single photons are extremely challenging, as the requirements for their implementation are very demanding and still have to find a viable solution~\cite{Hacker2016, javadi2015single, Heuck2020}.  
This motivates the search for alternative schemes that introduce nonlinearity in a more controllable and scalable manner.

In classical electronic hardware, nonlinearity and memory features are naturally combined within a single element, the memristor, a resistor whose resistance depends on the history of the applied current, thus incorporating  memory of past inputs~\cite{Chua1971, DiVentraPershin2023}. Memristors provide an efficient physical implementation of synaptic functionality and have become key components in the search and development of new neuromorphic computing architectures~\cite{Emboras2020,Kumar2022}. More broadly, physical systems combining nonlinear dynamics and memory are natural candidates for reservoir-computing approaches~\cite{lukovsevivcius2009reservoir,van2017advances,R_hm_2018,Tanaka_2019,yan2024emerging}, where a nonlinear dynamical system maps input data into a high-dimensional space, while only a simple readout layer is trained. This enables fast training and makes physical reservoirs particularly attractive, also in the quantum regime, where quantum reservoir computing~\cite{mujal2021opportunities,abbas2024classical} can be implemented by promoting the hidden layer to a quantum system. Therefore, motivated by the role of memristive dynamics in neuromorphic architectures, quantum memristors have been introduced as an architectural solution for neuromorphic quantum computation, providing a controlled way to realize effective nonlinearity~\cite{pfeiffer_quantum_2016}. The nonlinearity is introduced by correlating the evolution with its past measurement outcomes. At the same time, the dependence of its response on previous measurement outcomes gives rise to memory effects, thus enabling history-dependent input-output relations.

Quantum memristors have been investigated theoretically across multiple physical platforms, including superconducting circuits~\cite{salmilehto_quantum_2017}, trapped ions~\cite{stremoukhov_model_2024} and quantum dots~\cite{li_double_2017}, where they have also been experimentally explored. In parallel to these developments, photonic platforms have attracted growing interest due to their intrinsic advantages, such as low decoherence, room-temperature operation, and ultrahigh bandwidth, together with their natural compatibility with large-scale integrated photonic architectures~\cite{bao2023very,boes2018status,slussarenko2019photonic,kim2020hybrid}. These characteristics make photonics particularly well suited for implementing measurement-based adaptive dynamics and memory-dependent quantum processes~\cite{GAO2022100007, liu2025measurement}. 
Early photonic quantum memristors (PQMs) were based on a tunable beam splitter, whose reflectivity is adaptively updated according to measurement outcomes~\cite{sanz_invited_2018}. This scheme established the fundamental operating principle of PQMs in single-rail encoding.
Building upon this foundational concept, subsequent work introduced a dual-rail encoding approach~\cite{spagnolo2022experimental}, which permits the implementation of a photonic quantum memristive behavior using Mach–Zehnder interferometers (MZIs), a standard building block of photonic integrated circuits (PICs), thereby facilitating the extension of PQMs toward larger-scale implementations. The single-rail PQM has been implemented and extended to more complex architectures, such as cascaded PQMs~\cite{micco_quantum_2025}.
A single PQM~\cite{selimovic2025} as well as a feedback-driven linear circuit with multi-photon states~\cite{di2026time} have proven that reconfigurable linear optical circuits equipped with measurement-driven feedback laws are a valuable resource for quantum reservoir computing. In particular, these works have successfully solved temporal prediction tasks, such as temporal XOR, NARMA~\cite{ atiya2000new}, Mackey-Glass~\cite{MackeyGlass1977} and Santa Fe series~\cite{WeigendGershenfeld1993}.
Crucially, these implementations expand the prospects of investigating novel PQMs configurations and PICs with adaptive feedback schemes able to process information within the quantum neuromorphic paradigm. Indeed, on the one hand, improvements in PICs, such as low-loss waveguides with ultrahigh bandwidth, offer the opportunity to build large-scale and high-efficiency information processing units~\cite{luke2015broadband, wang2018integrated, he2019high}, while on the other hand, high-performance single photon sources exist, such as diamond color centers, that can offer near deterministic single photon emission at room temperature, alleviating experimental demands and enabling potential on-chip integration~\cite{lindner2018strongly,ngan2023quantum,fehler2020purcell}.

In this work, we report on the experimental investigation of two coupled PQMs with crossed feedback implemented on a silicon nitride PIC and fed by a single-photon source based on a SiV$^-$ color center in nanodiamonds operating at room-temperature. Silicon vacancy SiV$^-$ color centers feature a narrow emission peak (in contrast to nitrogen-vacancy centers NV$^-$) thanks to which they have already demonstrated their versatility across several quantum technologies, including solid-state quantum memories~\cite{berezhnoi2022quantum, pingault2017coherent, sukachev2017silicon} and elementary quantum networks~\cite{nguyen2019quantum}, where single photons emitted by SiV⁻ centers act as flying qubits interfacing distant spin-based nodes. Beyond networking, the availability of stable and spectrally narrow single-photon emission makes SiV-based sources potentially well-suited for linear optical quantum computing architectures, as there is still work to be done regarding integration, need for low temperature and multiple sources' indistinguishability. Using single photons from our nanodiamond-based single SiV$^-$ color center, we characterize both inter- and intra-memristor input-output relations, observing enhanced memory effects and more complicated effective nonlinear responses. Our results considerably extend the relevance of PQMs beyond isolated devices, enabling their exploration in networked architectures with complex connectivity and memory-induced entanglement dynamics~\cite{kumar_entangled_2021,kumar_tripartite_2022,ferrara_entanglement_2025}. In this context, suitably interconnected quantum memristors enabled by integrated photonics offer a physically grounded and scalable route toward quantum neural networks and photonic quantum simulators~\cite{selimovic2025,sakurai2025quantum,steinbrecher2019quantum}. This work constitutes the first implementation of PQMs combining memristive devices and a quantum light source both characterized by a high level of integrability, thus paving the way towards large-scale PQMs architectures.

The paper is structured as follows. Section~\ref{sec:source} introduces the single photons source. Section~\ref{sec:pic} describes the photonic implementation of a single PQM and two coupled PQMs. In Section~\ref{sec:results}, we present and discuss the experimental results. Finally, Section~\ref{sec:conclusions} draws the conclusions and reports a comparison with the other implementations of PQMs reported so far.

\section{Single photons source}
\label{sec:source}

To study the behaviour of coupled photonic quantum meristors, single photons from a negatively charged silicon vacancy SiV$^-$ color center were coupled to the input of the integrated circuit. Color centers are defects created or naturally present in the all-carbon lattice of diamond when adjacent carbon atoms are replaced by a vacancy-impurity atom combination that exhibits an atom-like electronic structure~\cite{Thiering2020_ColorCentersQuantum,Doherty2013_NVinDiamond}; in the case of SiV$^-$, this impurity is a Silicon atom. The so-called group-IV color centers, such as the SiV$^-$, exhibit bright, narrow-band, highly polarized and photostable single photon emission with short lifetimes~\cite{lagomarsino2021creation,Bra19}. In the SiV$^-$ color center, a silicon atom occupies an interstitial split-vacancy position between two adjacent empty lattice sites, giving the SiV$^-$ color center an inversion symmetry arising from the lack of a permanent electric dipole, which also shields the SiV$^-$ from electrical noise in its vicinity. The SiV$^-$ is a spin half system with ground and excited state levels that have partially lifted degeneracies due to spin-orbit coupling. The ground state sub-levels are separated by a splitting of 50 GHz while the excited state splitting is around 260 GHz. The zero phonon line (ZPL), which corresponds to the electronic transition between the excited and ground state, has a narrow peak at 737 nm in contrast to the better known nitrogen-vacancy center NV$^-$, which is characterized by a broad emission due to a strong phonon coupling~\cite{Doherty2013_NVinDiamond}.

\begin{figure}[!t]
    \centering
    \begin{subfigure}[b]{0.45\textwidth}
        \centering
        \includegraphics[width=\textwidth]{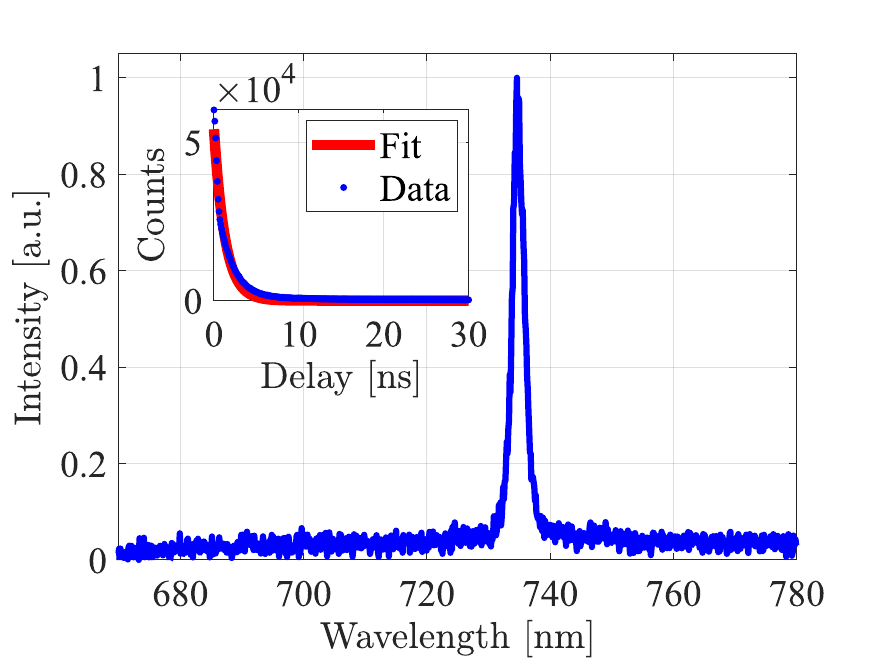}
        \subcaption*{(a)}
    \end{subfigure}
    \hfill
    \begin{subfigure}[b]{0.45\textwidth}
        \centering
        \includegraphics[width=\textwidth]{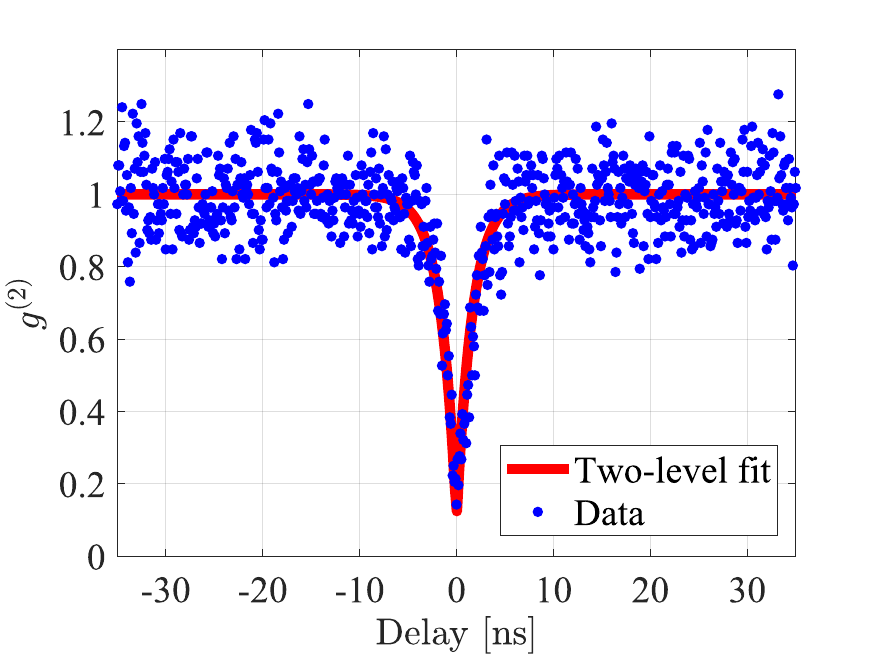}
        \subcaption*{(b)}
    \end{subfigure}    
    \caption{ (a) The photoluminescence spectrum of the SiV$^-$ showing a sharp ZPL peak at 734 nm. In the inset, the time decay of the photoluminescence (points) and its single exponential fit (line). The fit yields a lifetime of the SiV of 1.89 ns. (b) Second order correlation data (points) of the emitted photons as a function of the time delay between the two arms of the HBT interferometer. Data are fitted by a two levels model (line) which evidences a clear photon antibunching behaviour.  }
    \label{fig:source}
\end{figure}

The SiV$^-$ used in this work is hosted in a nanodiamond grown using a high pressure and high temperature (HPHT) process~\cite{lagomarsino2021creation}. This specific nanodiamond was selected from among those deposited on a silicon substrate via spin coating. The selection process was conducted using a homemade room temperature micro-photoluminescence ($\mu$-PL) setup by means of confocal intensity scans followed by photoluminescence measurements. For this specific SiV$^-$, we obtained a spectrum showing the ZPL to be at 734 nm as shown in Fig.~\ref{fig:source}(a). We attribute the deviation from the standard SiV$^-$ ZPL at 737 nm to the strain experienced by the color center within the nanodiamond lattice, which was previously reported in ~\cite{lindner2018strongly}. The $\mu$-PL setup used to characterize the color center is described in more detail in Appendix~\ref{app:setup}. The inset of Fig.~\ref{fig:source}(a) presents the time decay of the photoluminescence from which we estimated a lifetime of  $1.89 \pm 0.02$ ns. The single photon emission was confirmed using a Hanbury Brown and Twiss interferometer with the measurement of the photon autocorrelation function $g^{(2)}(0)$ showing an antibunching dip of $0.098 \pm 0.032$ and a FWHM (full width at half maximum) of $2.25 \pm 0.11$ ns. These data were obtained by a fit of the experimental points with a 2-level emitter model (red curve in Fig.~\ref{fig:source}(b)). The FWHM value is consistent with the measured lifetime. The $g^{(2)}(0)$ value is limited by the background light coming from the untreated surface of the nanodiamond. The $\mu$-PL optical setup was coupled to a single mode fiber yielding a brightness at the fiber output of 170 kcps. All measurements were initially performed at the University of Technology of Troyes (UTT), where the emitter position was recorded relative to the silicon substrate corners using a homemade triangulation protocol (see Appendix~\ref{app:Triangulation} for more details). This enabled the $\mu$-PL setup to be disassembled, transported to the University of Trento, and rapidly reassembled while reliably relocating the same emitter (see Appendix~\ref{app:setup}). As a result, the triangulation protocol effectively enabled the portability of the single-photon source.

\section{Integrated photonic quantum memristors}
\label{sec:pic}

The memristor concept was first proposed by Chua in the 1970s~\cite{Chua1971, ChuaKang1976} as a fundamental circuit element that relates magnetic flux and electric charge, as predicted by symmetry considerations in nonlinear electrical circuits. A memristor was demonstrated for the first time in 2008 with a nanoscale electronic circuit by Strukov et al.~\cite{Strukov2008}. The notion of memristive systems was soon extended to capacitive and inductive elements~\cite{DiVentraPershinChua2009}. Nowadays, it has been made clear that memelements are physical devices that satisfy specific physical properties~\cite{DiVentraPershin2023}. In a general form, the memristive behavior can be described by
\begin{equation}
y = f(s,x,t) x, \quad
\dot{s} = g(s,x,t),
\label{eq:mem_master_equation}
\end{equation}
where $x$ and $y$ denote the input and output variables, respectively, $s$ represents the internal state of the memristor, and $f$ and $g$ are functions that depend on the input, time, and the system’s previous states. This formulation explicitly captures both the nonlinear response and the memory effect of the device: the output $y$ depends on the input $x$ through a state-dependent function $f(s,x,t)$, while the evolution of the internal state $s$ is driven by the past and present values of the input, giving rise to history-dependent dynamics and hysteresis effects. For periodic input $x$, a pinched hysteresis loop in the input-output characteristic plane occurs, and the area of such a loop is associated with the memory~\cite{millar1951cxvi,biolek2014interpreting}. In order to compare memory effects despite of the maximal values of the characteristic input-output quantities or for systems affected by a dissipative dynamics~\cite{kumar_entangled_2021}, the size of the loop is quantified by a scaling-invariance parameter called form factor $F$, which is equal to $4\pi \, A/P^2$, where $A$ and $P$ are the area and the perimeter of the hysteresis curve, respectively.\\
With the aim of extending the concept of a classical memristor to the quantum regime, the most direct approach is to promote classical variables to quantum observables, while ensuring that the evolution remains coherent to preserve genuinely quantum features. 
However, the defining characteristics of a memristor—nonlinearity and memory effects—necessarily require non-Markovian dynamics, which cannot arise from closed, unitary evolution alone. 
This naturally calls for a description in terms of open quantum systems, where controlled measurement processes or environment-induced back-action provide a physically consistent mechanism to generate history-dependent and nonlinear responses. 

\begin{figure}[!t]
    \centering
    \begin{subfigure}[b]{0.23\textwidth}
       \centering
        \includegraphics[width=1.4\textwidth]{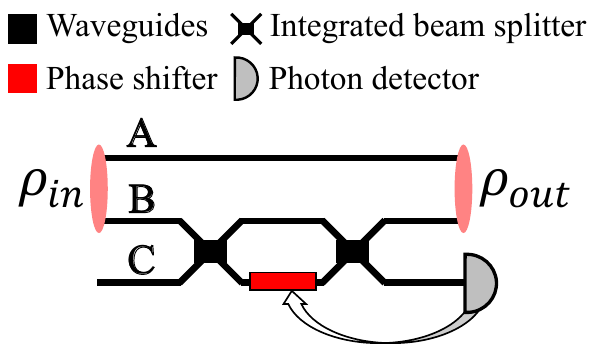}
        \subcaption*{(a)}
    \end{subfigure}
    \hfill
    \begin{subfigure}[b]{0.7\textwidth}
        \centering
        \includegraphics[width=0.95\textwidth]{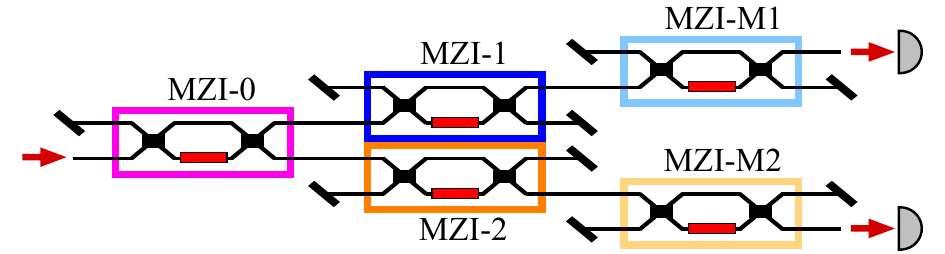}
        \subcaption*{(b)}
    \end{subfigure}
    \caption{ (a) Dual-rail scheme of the photonic quantum memristor, where the input state $\rho_{in}$ is encoded as a single photon in a superposition of the upper path modes A-B. The component in mode B is transformed by the Mach–Zehnder interferometer, implementing the photonic quantum memristor, while the component in mode A propagates freely to the output state $\rho_{out}$. The Mach-Zehnder interferometer is composed of two beam splitters (black rectangles), and a thermal phase shifter (red rectangle). The output on mode C is used as a feedback for the phase setting of the memristor. (b) Schematic layout of the circuit implementing a single (MZI-M1) and two coupled integrated photonic quantum memristors (MZI-M1 and MZI-M2)). The red arrow on the left indicates the input waveguide, while the arrows on the right are placed to highlight the output waveguides, fiber-coupled to single-photon SPAD detectors, which are represented by grey semi-circles. MZI-0, MZI-1, MZI-2 are used to prepare the suitable input state for the PQM. The input and output waveguides that are not used in the experiments are pointed out by a black oblique line. }
    \label{fig:pic_scheme}
\end{figure}

In photonic platforms, a PQM, can be implemented using a beam splitter~\cite{sanz_invited_2018} or an MZI~\cite{spagnolo2022experimental}, which are denoted as single- and dual-rail devices, respectively. For a single PQM, these two schemes are equivalent. In the latter scheme, as shown in Fig.~\ref{fig:pic_scheme}(a), 
a photonic qubit is encoded in the path degree of freedom of the two upper modes A and B. 
The input state is in superposition:  
\begin{equation}
    |\psi_{in}(t)\rangle\equiv\alpha(t) |1_{\rm A},0_{\rm B}\rangle + \beta(t) |0_{\rm A},1_{\rm B}\rangle \,,
    \label{eq:initialstate_singlePQM}
\end{equation}
and the input density matrix is $\rho_{in}(t)\equiv|\psi_{in}(t)\rangle\langle\psi_{in}(t)|$, where the qubit state $|0_{\rm A/B}\rangle$ corresponds to zero photons in mode A/B and $|1_{\rm A/B}\rangle$ corresponds to a photon in mode A/B. As usual, $|\alpha(t)|^2+|\beta(t)|^2=1$.
Then, an auxiliary mode C without photons is added to the input state, giving the total state $|\psi_{in}(t)\rangle\otimes|0_{\rm C}\rangle$, and the transformation $U(t)$ associated to the MZI acting on modes B and C is applied, i.e. $U(t) \,\rho_{in}(t) \otimes|0_{\rm C}\rangle\langle0_{\rm C}|\, U(t)^\dagger$.
Given the input state and denoting the reflectance of the MZI as $R(t)$, the MZI map reads: 
\begin{equation}
\begin{split}
     U(t) &= \mathbf{1}_{\rm A}\otimes \left[|0_{\rm B},0_{\rm C}\rangle\langle 0_{\rm B},0_{\rm C}|+\sqrt{1-R(t)}\,|1_{\rm B},0_{\rm C}\rangle\langle 1_{\rm B},0_{\rm C}|+\sqrt{R(t)}\,|0_{\rm B},1_{\rm C}\rangle\langle 1_{\rm B},0_{\rm C}| \right. \\
     & \hspace{2cm}\left. +\sqrt{1-R(t)}\,|0_{\rm B},1_{\rm C}\rangle\langle 0_{\rm B},1_{\rm C}|-\sqrt{R(t)}\,|1_{\rm B},0_{\rm C}\rangle\langle 0_{\rm B},1_{\rm C}| \right]\,, 
\end{split}
\end{equation}
where $\mathbf{1}_{\rm A}$ is the identity map over the mode A. Note that our convention implies that a ``transmitted photon'' is a photon entering in mode B/C and remaining in mode B/C after the MZI with probability $1-R(t)$, while a ``reflected photon'' is a photon entering in mode B/C and exiting in mode C/B with probability $R(t)$.
The output density matrix $\rho_{out}(t)$ is defined as the partial trace over the mode C of the density matrix $U(t) \,\rho_{in}(t) \otimes|0_{\rm C}\rangle\langle0_{\rm C}|\, U(t)^\dagger$. This means that 
\begin{equation}
\begin{split}
    \rho_{out}(t) &\equiv \langle0_{\rm C} |U(t) \,\rho_{in}(t) \otimes|0_{\rm C}\rangle\langle0_{\rm C}|\, U(t)^\dagger|0_{\rm C}\rangle+\langle 1_{\rm C} |U(t) \,\rho_{in}(t) \otimes|0_{\rm C}\rangle\langle0_{\rm C}|\, U(t)^\dagger|1_{\rm C}\rangle 
    \\
    & = \left( \alpha(t)|1_{\rm A},0_{\rm B}\rangle +\beta(t) \sqrt{1-R(t)}|0_{\rm A},1_{\rm B}\rangle\right)\left( \bar{\alpha}(t)\langle1_{\rm A},0_{\rm B}| +\bar{\beta}(t) \sqrt{1-R(t)}\langle0_{\rm A},1_{\rm B}| \right) \\
    & \qquad + |\beta(t)|^2 R(t) |0_{\rm A},0_{\rm B}\rangle\langle0_{\rm A},0_{\rm B}|\,.
\end{split}
\end{equation}
The average input photon number in mode B is $\langle N_{in}(t)\rangle \equiv {\rm Tr}\left[ \rho_{in} \mathbf{1}_{\rm A}\otimes|1\rangle_{\rm B}\langle 1|_{\rm B}\right] =|\beta(t)|^2 
$ and the corresponding output average photon number 
$\langle N_{out}(t) \rangle \equiv {\rm Tr}\left[ \rho_{out} \mathbf{1}_{\rm A}\otimes|1\rangle_{\rm B}\langle 1|_{\rm B}\right] = [1 - R(t)] \langle N_{in}(t) \rangle$.
Finally, the quantum memristor is implemented by an MZI acting on mode B and an auxiliary mode C through the tunable beam-splitting reflectance $R(t)$ of the MZI, which is related to the phase setting of the MZI. Thus, looking at Eq.~\eqref{eq:mem_master_equation}, $(\langle N_{in}\rangle,\langle N_{out}\rangle)$ correspond to $(x,y)$, and $R$ plays the role of the internal memristive state $s$, which is controlled by previous measurement outcomes. This measurement process effectively couples the device to an environment and renders the dynamics non-unitary. In fact, the measurement outcomes are used to update the internal state variable $R(t)$, such that its evolution depends on the past output of the system. As a consequence, the current input–output relation is conditioned on the system’s history, realizing a genuinely non-Markovian dynamics.
Since the component in mode A propagates freely to the output state $\rho_{out}$ and only the component in mode B participates in the memristive feedback dynamics, 
the evolution of the single quantum memristor can be fully characterized by the dynamics of the measurable input and output average photon numbers, 
$\langle N_{in}(t) \rangle$ and $\langle N_{out}(t) \rangle$. These observables can be evaluated from the photon flows on output spatial modes A, B and C. In particular, $\langle N_{in}(t) \rangle$ can be found from the average photon flow on output mode C divided by $R(t)$. 
The quantum coherence $C^{\rm single}$ of the single PQM~\cite{ferrara_entanglement_2025} is related to the moduli of the off-diagonal terms of the density matrix~\cite{Baumgratz_2014}:
\begin{equation}
    C^{\rm single}_{in/out} = \sum_{i\ne j} |\langle i| \rho_{in/out} (t) | j\rangle | = 2 |\beta(t)| \sqrt{1-|\beta(t)|^2} \times\begin{cases}
         1 & \mbox{if input}\,, \\
         \sqrt{1-R(t)} & \mbox{if output} \,.
    \end{cases}
\end{equation}
Both input and output coherences vary in time modulo a factor $\sqrt{1-R(t)}$: this means that quantum coherence is maintained despite of the measurement and the tracing out of mode C. The von Neumann entropy~\cite{nielsen_chuang_2010} of the single PQM output is not zero contrary to its input.\\
To illustrate the basic operating principle, the nonlinearity and memory effects of the memristor can be probed by observing the emergence of a hysteresis loop~\cite{sanz_invited_2018}.
This requires a cyclic driving protocol in which the input variable is swept from small to large values and back again, such that the output depends not only on the instantaneous input but also on its history. For this reason, the input photon number $\langle N_{in}(t) \rangle$ is chosen to vary sinusoidally between zero and a maximum value $\langle N_{max} \rangle$ according to $\langle N_{in}(t)\rangle=\sin^2(\pi\,t/T_{\rm osc})$, where $T_{\rm osc}$ is the oscillation period and all photon numbers are normalized by the maximum value $\langle N_{max} \rangle$.
The reflectance $R(t)$ is updated by the following relation: $\dot{R}(t) = \langle N_{in}(t) \rangle-0.5$, similarly to Eq.~\eqref{eq:mem_master_equation}. The resulting dynamics of $R(t)$ can then be written as~\cite{spagnolo2022experimental}:
\begin{equation}
    R(t)= \frac{1}{2}+\frac{1}{T}\int_{t-T}^{t}\left(\langle N_{in}(t')\rangle-\frac{1}{2}\right) dt' \,.
    \label{eq:feed_law_cont}
\end{equation}
Here, the integration window $T$ defines the memory depth of the quantum memristor, characterizing how many past input states contribute to the current internal state.
Our experiments with the single memristor and the coupled memristors are performed in $N$ time bins of temporal duration $\tau$. Starting from the time bin duration, the characteristic times are $T_{\rm osc}=M_{\rm osc} \tau$ and $T=M \tau$, with $(M_{\rm osc}, M)$ being integers. The first one is related to the period of the sinusoidal oscillation of the input photon flux, while the second one is related to the buffer length of the memristor. Each time bin $t_k$ is associated to an input $\langle N_{in} (t_k)\rangle$ evaluated from the number of detected photons on output modes B and C divided by the total number of detected photons on all the output modes. Analogously, the output $\langle N_{out} (t_k)\rangle$ is found from the number of detected photons on output modes B divided by the total number of detected photons on all the output modes.
The feedback law in Eq.~\eqref{eq:feed_law_cont} becomes
\begin{equation}
R(t_k) = \frac{1}{2} + \frac{1}{M} \sum_{j=k-M+1}^k \left( \langle N_{in} (t_j)\rangle -\frac{1}{2} \right) \,.
\label{eq:feed_law_disc}
\end{equation}
$R(t_k)$ is then translated in a specific  phase setting of the MZI implementing the PQM during the time bin $t_k$.
In our experiments, we chose $\tau=20$ s plus the dead times due to electrical communications and phase settings of the MZIs. This time allows to collect enough detector's counts in order to estimate the observables $\langle N_{in} (t_k)\rangle$ and $\langle N_{out} (t_k)\rangle$.
Importantly, the integrated photonics paradigm makes this approach especially attractive, as MZIs are standard and highly controllable building blocks of PICs, enabling compact, stable, and reproducible implementations. In the present setup, however, the single-emitter SiV nanodiamond source remains off-chip: photons are collected using a confocal $\mu$PL setup coupled to a tapered-lens fiber, whose coupling to the PIC hosting the memristive circuit introduces additional losses. In this context, on-chip single-photon sources would represent a key step toward scalability by reducing coupling losses and facilitating the transition from isolated quantum memristive elements to larger networks of coupled quantum memristors.

The PIC used for this work is contained in a silicon nitride chip, and the waveguides are designed to be single-mode (cross-section 550 x 150 nm$^2$) at the emission wavelength of the source, i.e. $\sim 734$ nm.  Multimode-interference-based integrated beam splitters (MMIs)~\cite{Soldanoilgrande} are optimized to work with the fundamental transverse electric field (TE$_0$). 
The integrated MZIs are made of two MMIs and a pair of thermal phase shifters, one above each waveguide of the MZI's arms~\cite{Harris_14}.
Fig.~\ref{fig:pic_scheme}(b) shows the schematic PIC layout. The circuit is composed of two sections: a triangular scheme of three MZIs, denoted as MZI-0, MZI-1 and MZI-2, and two parallel MZIs denoted as MZI-M1 and MZI-M2.
We inject the single photons into the second input waveguide of MZI-0 (counting from the top, see red arrow). Then, only one output of MZI-M1 and one output of MZI-M2 (see red arrows in front the detectors) are connected to fiber-coupled single-photon silicon avalanche diode (SPAD) detectors. The choice of the output channels is related to the properties of the MZIs: the transmission on the two outputs of MZIs is related by a $\pi/2$ phase shift, thus we can effectively sample both outputs by collecting photons at the same output channel with two different phase settings and two runs. This implies the doubling of the time of the experiments, but it allows the use of fewer detectors and outputs, which are generally characterized by different coupling losses. \\
In the case of the single-memristor experiments, MZI-0 is used to modulate the photon flux to the upper output with a sinusoidal shape in time, while MZI-1 and MZI-2 are set to transmit all the light to the upper and lower outputs, respectively. MZI-M1 is the integrated memristor, while MZI-M2 is used in a fixed setting to transmit all the photons on its lower output to compute the average photon numbers. 
In the coupled-memristor experiments, MZI-M1 and MZI-M2 serve as the integrated memristors working in parallel with independent input modulations. MZI-0 evenly and coherently splits the input flow between its upper and lower outputs, while MZI-1 and MZI-2 modulate the photon flux entering the two memristors with a sinusoidal time dependence.
We choose the buffer length of both memristors to be equal to $M$, and we introduce a relative phase shift $\Phi$ between the time modulation of MZI-1 and MZI-2. Only two detectors are used in this case as well, by exploiting the $\pi/2$ phase shift between the outputs of each MZI (see Appendix~\ref{app:coupled_mems} for details).
The feedback laws for MZI-M1 and MZI-M2 are given by: 
\begin{equation}
\begin{split}
& R^{(1)}(t_k) = \frac{1}{2} + \frac{1}{M} \!\!\sum_{j=k-M+1}^k \!\!\left( \langle N_{in}^{(2)}(t_j) \rangle -\frac{1}{2} \right)
\;\;,\;\;
 R^{(2)}(t_k) = \frac{1}{2} + \frac{1}{M} \!\!\sum_{j=k-M+1}^k \!\!\left( \langle N_{in}^{(1)}(t_j) \rangle -\frac{1}{2} \right)\,,
\end{split}
\label{eq:nest_feed_law}
\end{equation}
where the superscript refers to the first or second memristor. Thus, the setting of one memristor depends on the history of the other, and vice versa.
In analogy with the single PQM, we consider the observables $(\langle N_{in}^{(1)} \rangle,\langle N_{out}^{(1)}\rangle,\langle N_{in}^{(2)} \rangle,\langle N_{out}^{(2)} \rangle)$, i.e. the input and output average photon numbers of the two PQMs in parallel. However, since MZI-0 acts as a balanced beam-splitter for an input photon, the overall state is composed of terms describing the evolution of the individual PQMs plus an interference term between the two memristors' evolutions. This additional term gives rise to additional correlations. In our present geometry, we cannot exploit them, since the outputs of the two PQMs are not connected by a ``mixing'' device. In a future design, one can consider to insert an additional balanced beam-splitter to measure these correlations. 
In the present scheme, this fact means that the source can be replaced by an attenuated coherent source. 
We refer to Appendix~\ref{app:quantumcoupledPQMs} for a detailed derivation.

\section{Results and discussions}
\label{sec:results}

\begin{figure}[!t]
  \centering
  \includegraphics[width=\linewidth]{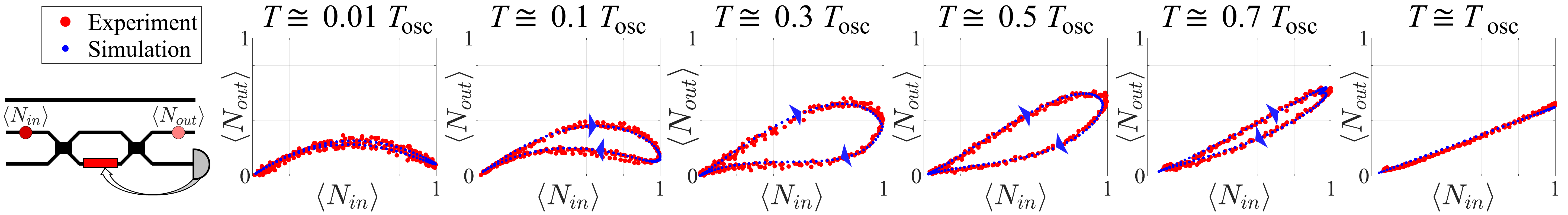}
  \caption{ Experimental and simulation results for the hysteresis loop of a single photonic quantum memristor with different ratios between the two time scales: oscillation period of the modulated input flux, $T_{\rm osc}$, and time length of the memristor's buffer, $T$. On the left, a schematic diagram of the device is shown. For non-zero area, blue triangles show the direction of the hysteresis loops with respect to the sweep of the input photon number. Being small, error bars are neglected here for clarity. The raw data are affected by 1-5$\%$ relative error.}
  \label{fig:single_mem}
\end{figure}

Using the single-photon source described in Section~\ref{sec:source} and the PIC presented in Section~\ref{sec:pic}~\cite{baldazzi2024linear}, the experiments with a single PQM and two parallel PQMs with crossed feedback laws are executed for different configurations. In particular, for both cases the ratio between $T_{\rm osc}$ and $T$ is varied, while in the second case also the phase shift between the input single photon fluxes is changed.
Before every measurement, we run a calibration routine of all the used integrated devices, namely the MZIs highlighted in Fig.~\ref{fig:pic_scheme}(b). The calibration gives not only the relation between the phase and the electrical power of each thermal phase shifter, but also the interference visibilities of the associated MZI. Due to the  insertion losses of the PIC (20dB) and the single-photon rate of the source (170 kcps), in-fiber polarization control of the input signal is not feasible. In addition, alignment of the single-photon source is performed by adjusting a non-polarization-maintaining output fiber and directly coupling it to the SPAD detectors. As a result, the polarization state of the injected photons can change after each realignment, requiring recalibration of the MZIs, which are polarization-sensitive. Indeed, as they are designed to operate in the TE$_0$ mode, any TM$_0$ component in the guided light reduces the visibility of the interference they produce.
Thus, there are systematic errors of the MZIs given by the polarization in input and stochastic errors given by polarization drifts, which affect the visibility and current-phase relation of the PSs.

\begin{figure}[!t]
    \centering
    \begin{subfigure}[b]{0.49\textwidth}
        \centering
        \includegraphics[width=\textwidth]{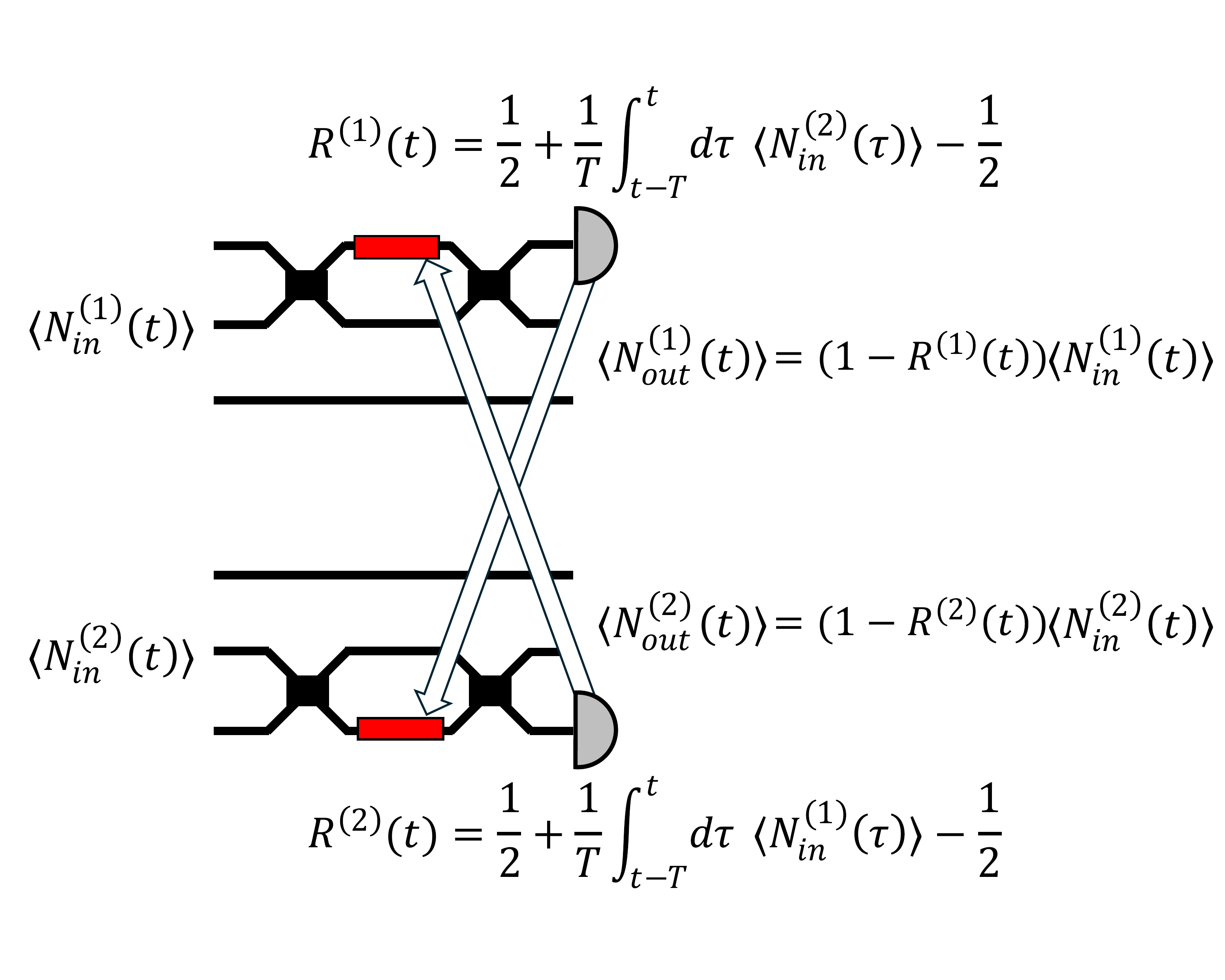}
        \subcaption*{(a)}
    \end{subfigure}
    \hfill
    \begin{subfigure}[b]{0.49\textwidth}
        \centering
        \includegraphics[width=\textwidth]{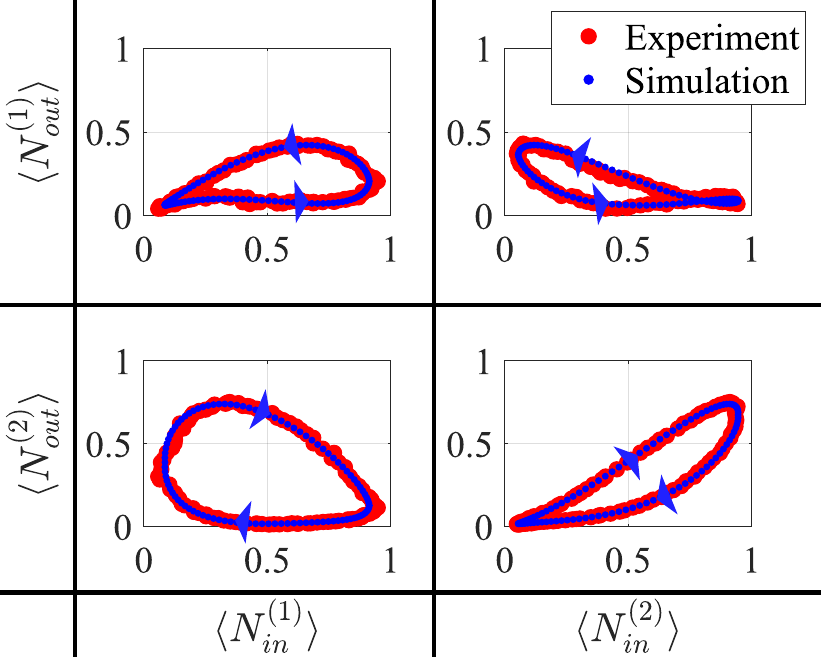}
        \subcaption*{(b)}
    \end{subfigure}
    \hfill
    \begin{subfigure}[b]{0.49\textwidth}
        \centering
        \includegraphics[width=\textwidth]{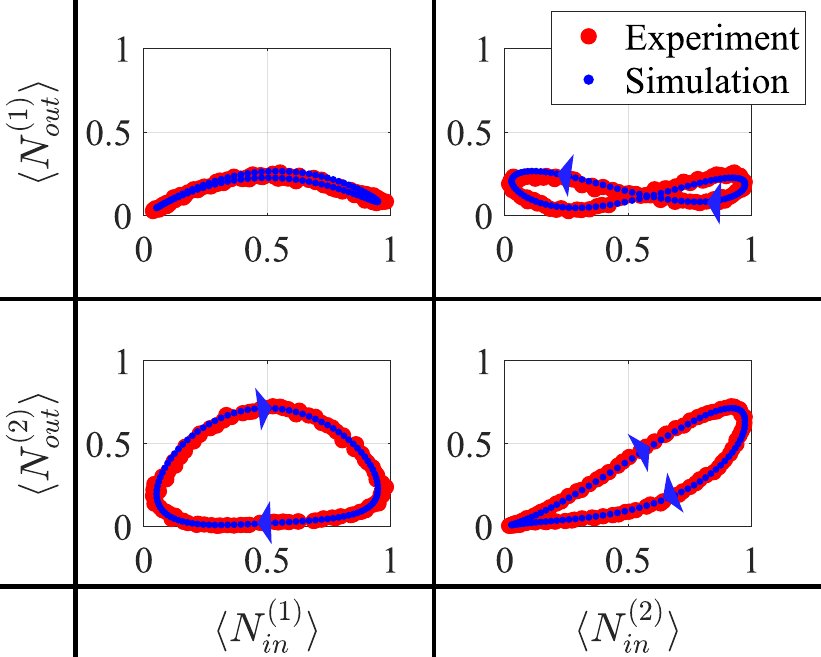}
        \subcaption*{(c)}
    \end{subfigure}
    \hfill
    \begin{subfigure}[b]{0.49\textwidth}
        \centering
        \includegraphics[width=\textwidth]{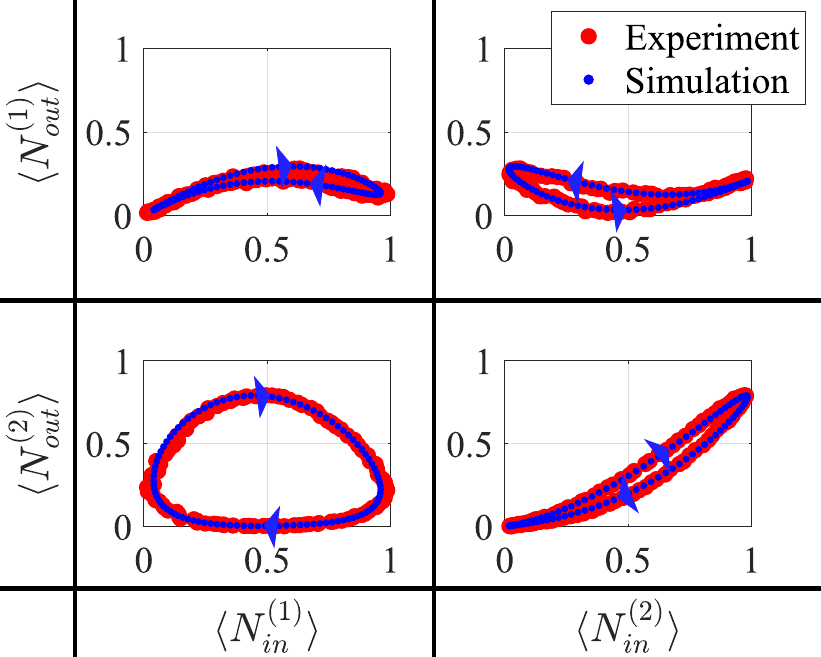}
        \subcaption*{(d)}
    \end{subfigure}
    \caption{(a) Schematic representation of the coupled photonic memristors with crossed feedbacks whose relations are described in integral form. Eq.~\eqref{eq:nest_feed_law} shows the corresponding discrete version.
    (b-d) Experimental and simulation results of coupled photonic quantum memristors for different ratios $T/T_{\rm osc}$ and input phase difference $\Phi$. Both single photonic quantum memristors receive a sinusoidal input with the same period $T_{\rm osc}$ and they have the same buffer length $T$. In particular, (b) $T=0.2\,T_{\rm osc}$ and $\Phi=0.7$rad. (c) $T=0.3\,T_{\rm osc}$ and $\Phi=0.5$rad. (d) $T=0.4\,T_{\rm osc}$ and $\Phi=0.7$rad. The four panels refer to the intra (diagonal cells) and inter (off-diagonal cells) memristor relations.
    For non-zero area, blue triangles show the direction of the hysteresis loops with respect to the sweep of the inputs photon number. Experimental errors are smaller than the data markers. The data are affected by 1-5$\%$ relative error.}
    \label{fig:coupled_mems}
\end{figure}

Fig.~\ref{fig:single_mem} presents the results for a single PQM. We set $T_{\rm osc} =100\,\tau$. Depending on the ratio between the two characteristic times, $T_{\rm osc}$ and $T$, different input-output curves arise. When $T=0.01\, T_{\rm osc}=\tau$, the average input flux $\langle N_{in}\rangle$ at previous times does not enter in the feedback law, and the input-output relation is almost a parabola, first curve in Fig.~\ref{fig:single_mem}. For $\tau<T<T_{\rm osc}$ (second to fifth curves in Fig.~\ref{fig:single_mem}), the input-output curve is characterized by a hysteresis figure pinched at the origin \footnote{A hysteresis figure is pinched at a point when its ascending and descending branches are tangent to each other in this point~\cite{DiVentraPershin2023}.}, while remaining nonlinear. The curves are taken with clockwise orientation for a periodic sweep of the input photon number.
Finally, for $T=T_{\rm osc}$ the input-output relation becomes temporally punctual (memory effect is lost) and linear, last curve in Fig.~\ref{fig:single_mem}. These behaviours are analogous to the electric case~\cite{Chua1971,Strukov2008}, since the feedback law considers the average of the input flux over a full period. 
Together with the experimental data, numerical simulation results are also reported (the blue curves). In the simulations results reported in Fig.~\ref{fig:single_mem} (see Appendix~\ref{app:simulations} for details), the non-unitary visibilities extracted from the calibration routine are included for all the MZIs used in the experiments. 

\begin{figure}[!t]
  \centering
  \includegraphics[width=\linewidth]{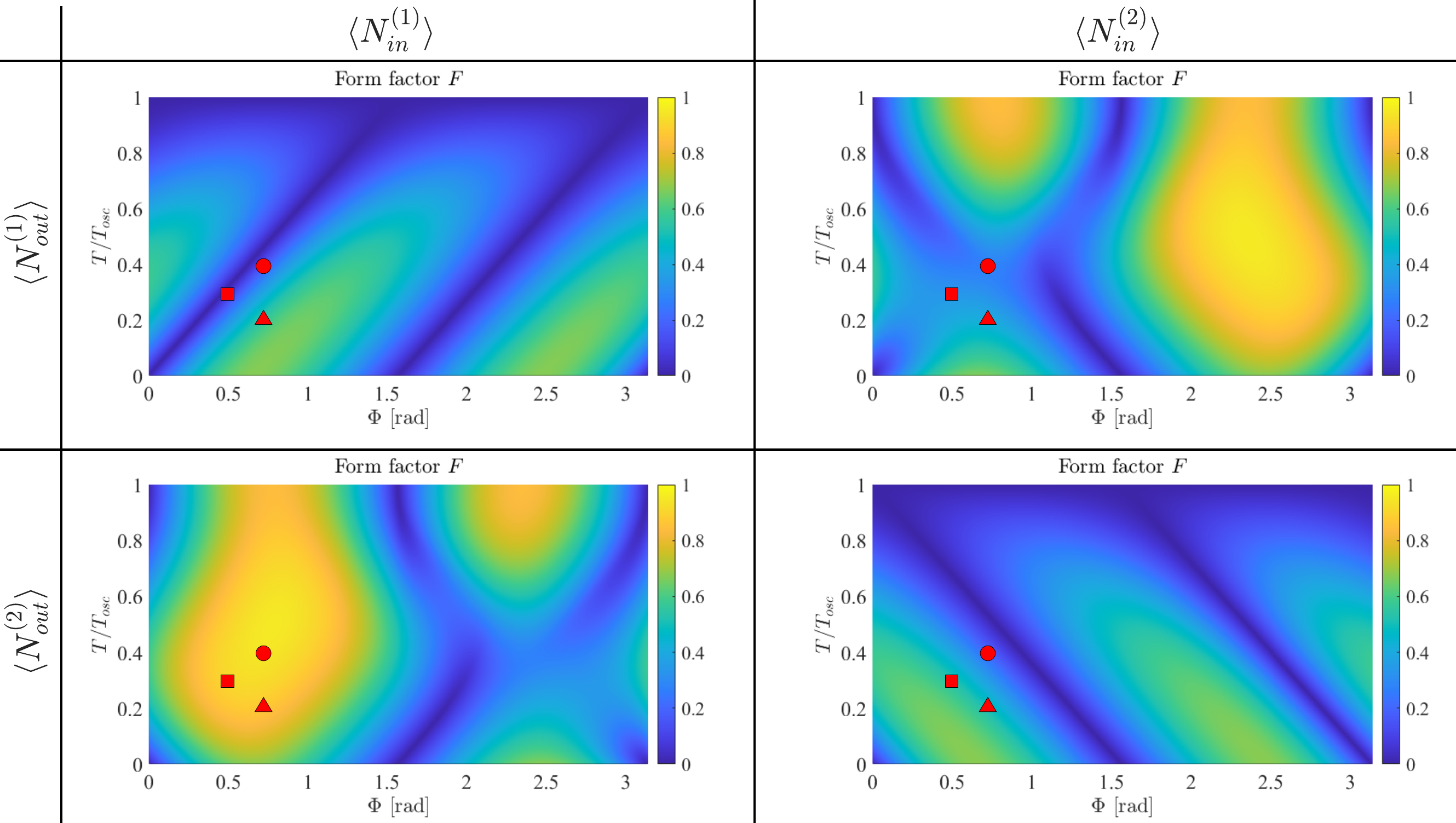}
  \caption{Simulations of the form factor $F$ of the intra- and inter-memristor hysteresis curves of two coupled photonic quantum memristors with crossed feedbacks as a function of the relative phase $\Phi$ between their inputs and the ratio $T/T_{\rm osc}$ between the buffer length of the memristors, $T$, and the period of the sinusoidal input modulation, $T_{\rm osc}$. The form factor $F$ is $4\pi$ times the ratio between the area and the squared perimeter of the hysteresis curve. The intra-relations, $(\langle N_{in}^{(1)} \rangle,\langle N_{out}^{(1)} \rangle)$ and $(\langle N_{in}^{(2)} \rangle,\langle N_{out}^{(2)} \rangle)$, are on the diagonal of the table, and the inter-relations, $(\langle N_{in}^{(2)} \rangle,\langle N_{out}^{(1)} \rangle)$ and $(\langle N_{in}^{(1)} \rangle,\langle N_{out}^{(2)} \rangle)$, are on the off-diagonal. The values have been normalized with respect to the maximum value of $F$ for the inter-relations, i.e. $\approx 0.95$ for $T/T_{\rm osc}\approx 0.5$ and relative phases $\Phi\approx 2.35$ rad and $\Phi\approx 0.77$ rad, respectively. The red symbols in the surface plots indicate the choice of parameters used in the experiments of the two coupled memristors with crossed feedback reported in Fig.~\ref{fig:coupled_mems}: in particular, the triangle-square-circle represents the choice associated with the hysteresis curves in Fig.~\ref{fig:coupled_mems}(b-c-d), respectively.}
  \label{fig:simu_nested_mems_form_factor}
\end{figure}

In the case of the coupled PQMs, we choose the times $(T_{\rm osc}, T)$ of both devices to be equal, and we set $T_{\rm osc} =100\,\tau$, as in the single-memristor case. Within this configuration, we studied the dependence of the input-output relations with respect to the ratio $T/T_{\rm osc}$ and to the relative phase $\Phi$ between the sinusoidal inputs. The outcome is simply the one of two independent PQMs for zero phase difference, while non-trivial intra- and inter-memristor relations between the input-output fluxes arise for a non-zero phase difference. Here and in the following, we refer to  $(\langle N_{in}^{(i)} \rangle,\langle N_{out}^{(i)} \rangle)$ as intra-relations,  with $i=\{1,2\}$, and to  $(\langle N_{in}^{(i)} \rangle,\langle N_{out}^{(j)} \rangle)$ as inter-relations,  with $i,j=\{1,2\}$ and $i\neq j$.
Fig.~\ref{fig:coupled_mems} shows the experimental results, together with simulations with the real visibilities, for different configurations of inputs' phase difference $\Phi$ and ratio $T/T_{\rm osc}$. Appendix~\ref{app:simulations} contains the parameters of the simulations.
We observe that peculiar hysteresis figures can be obtained for inter-relations, sharing the common feature of not being pinched at the origin. In particular, Figs.~\ref{fig:coupled_mems}(b-c-d) show 'large area' curves for $(\langle N_{in}^{(1)} \rangle,\langle N_{out}^{(2)} \rangle)$ and 'infinity loop'-like self-intersecting curve for $(\langle N_{in}^{(2)} \rangle,\langle N_{out}^{(1)} \rangle)$. The former type of hysteresis curve is interestingly characterized by the fact of featuring a marked bistable behaviour. While the second aspect, to the best of our knowledge, has not been reported for electrical memelements yet: a self-intersection point implies that the derivative of the curve is not unique.
These features introduce more complexity in the system and agree with former observations reporting that non-pinched hysteresis loop under periodic drive can be found e.g. in several electrical devices with memory~\cite{DiVentraPershin2023}. 

\begin{figure}[!t]
  \centering
  \includegraphics[width=\linewidth]{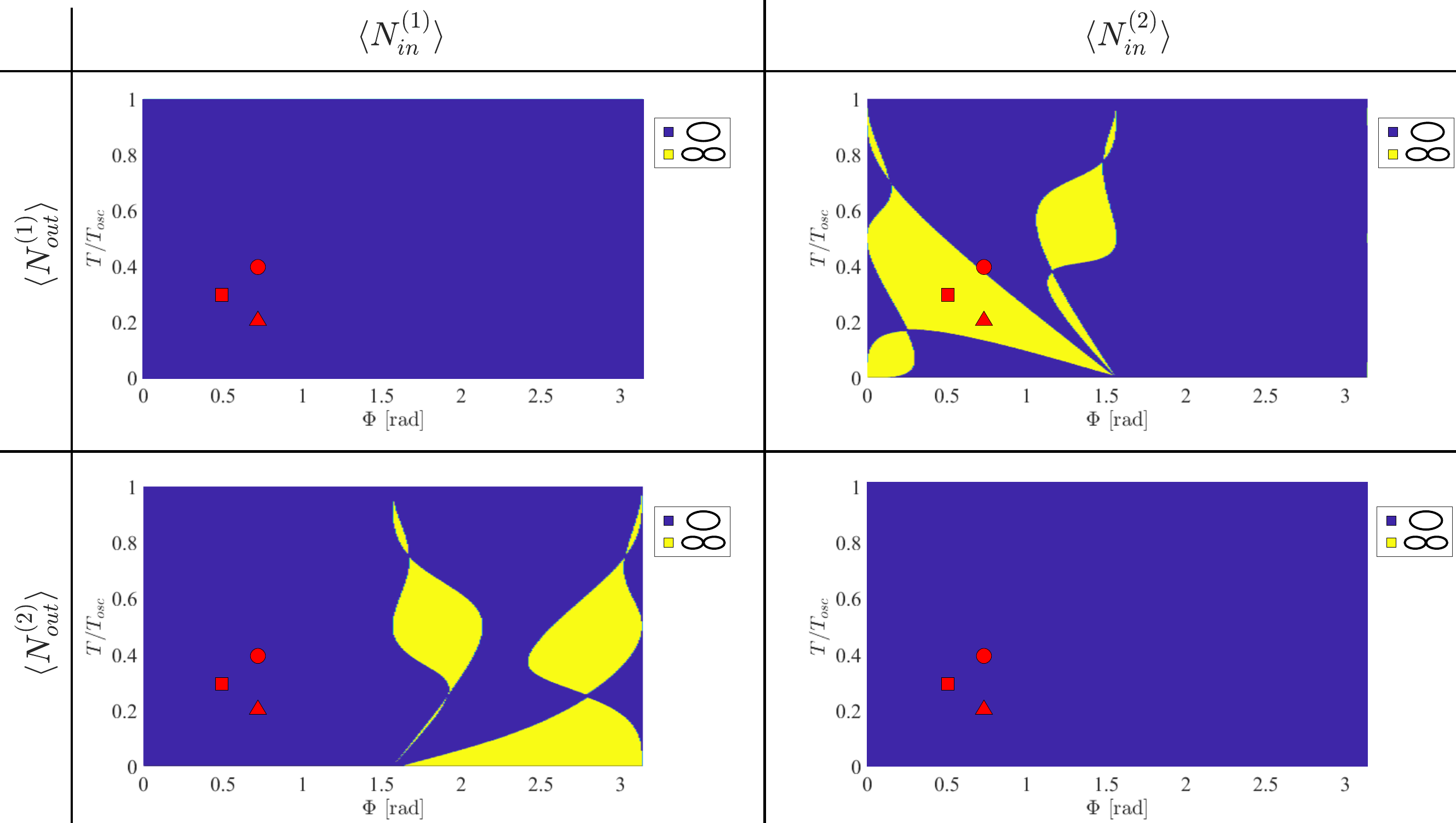}
  \caption{Simulations results revealing the presence (yellow regions) of a self-intersecting hysteresis loop in two coupled PQMs with crossed feedback as a function of the relative phase $\Phi$ between the inputs and the ratio $T/T_{\rm osc}$ between the buffer length of the memristors, $T$, and the period of the sinusoidal input modulation, $T_{\rm osc}$. The infinity symbols and the yellow stands for self-intersecting curves, while the ellipse and the blue for not-self-intersecting curves. The red symbols in the surface plots indicate the choice of parameters used in the experiments of the two coupled memristors with crossed feedback reported in Fig.~\ref{fig:coupled_mems}: in particular, the triangle-square-circle represents the choice associated with the hysteresis curves in Fig.~\ref{fig:coupled_mems}(b-c-d), respectively.} 
  \label{fig:simu_nested_mems_self_inter}
\end{figure}

Relevant features of the hysteresis curves of coupled PQMs are the form factor and the presence of a point of self-intersection. 
We investigated both features in numerical simulations by varying the relative phase $\Phi$ between the input fluxes and the ratio $T/T_{\rm osc}$ between the buffer length of the memristors and the period of the sinusoidal modulation of the inputs. Fig.~\ref{fig:simu_nested_mems_form_factor} shows the form factor for the four hysteresis curves $(\langle N_{in}^{(1/2)} \rangle,\langle N_{out}^{(1/2)} \rangle)$ with different choices of $\Phi$ and $T/T_{\rm osc}$. 
For the intra-relations, there are constant-level lines of the form factor. The points indicated with the red square and circle/triangle in  Fig.~\ref{fig:simu_nested_mems_form_factor} share the same value of form factor: this is manifest by looking at the corresponding intra-relations in Fig.~\ref{fig:coupled_mems}(c) and Fig.~\ref{fig:coupled_mems}(d)/(b), respectively.
Very interestingly, the maxima of the form factor of the hysteresis curves of intra-relations, namely $(\langle N_{in}^{(1)} \rangle,\langle N_{out}^{(1)} \rangle)$ and $(\langle N_{in}^{(2)} \rangle,\langle N_{out}^{(2)} \rangle)$, have value $\approx 0.67$ for $T/T_{\rm osc} \ll 1$ and relative phases equal to $\Phi\approx(0.65,2.47)$ rad. So, even if the single memristors have a buffer with only one element, the largest form factor arises from the correlations due to the crossed feedback, Eq.~\eqref{eq:nest_feed_law}, and specific values of phase shift $\Phi$ between the input single photon fluxes. The form factors of the inter-relations, $(\langle N_{in}^{(2)} \rangle,\langle N_{out}^{(1)} \rangle)$ and $(\langle N_{in}^{(1)} \rangle,\langle N_{out}^{(2)} \rangle)$, have maximum of $\approx 0.95$ for $T/T_{\rm osc}\approx 0.5$ and relative phases $\Phi\approx 2.35$ rad and $\Phi\approx 0.77$ rad, respectively. Thus, the form factor of the intra- and inter-relations has a maximum higher than the form factor obtained with a single memristor, i.e. $F\approx0.58$ for $T/T_{\rm osc} \approx 0.35$, which is a clear indication of enhanced bistable behaviour.
Fig.~\ref{fig:simu_nested_mems_self_inter} reports the map of the regions where a self intersection point can be observed in the hysteresis curves as a function of $\Phi$ and $T/T_{\rm osc}$. Note that there are no self-intersecting curves for the intra-relations, while several choices of $\Phi$ and $T/T_{\rm osc}$ produce a self-intersecting point for only one inter-relation curve. Indeed, a self-intersection point for one inter-relation curve excludes a self-intersection point for the other inter-relation curve. 

In numerical simulations, we have also investigated the case of two individual PQMs with the same non-crossed feedback law: the intra-relations trivially have the maximum value of the form factor equal to the maximum of the single PQMs, and they show no intersection points; the inter-relations are characterized by a maximum value of the form factor equal to $\approx 0.87$, lower than the crossed-feedback case, and they show self-intersection points. However, contrary to the crossed-feedback scenario, both inter-relations feature a self-intersecting point for the same values of $\Phi$ and $T/T_{\rm osc}$. Therefore, we conclude that the property of self-intersection for inter-relations is originated from the phase relation between the two periodic inputs, but the exclusivity of having a self-intersection point for only one inter-relation and not the other comes from the crossed feedback law, Eq.~\eqref{eq:nest_feed_law}.
Further investigations are needed to explore the parameter configuration of two memristors in parallel: for example, the buffer lengths of the two memristors can be chosen with different values and the feedback laws can be taken as a convex combination of the non-crossed and crossed feedback laws.

\begin{figure}[!t]
    \centering
    \begin{subfigure}[b]{0.48\textwidth}
        \centering
        \includegraphics[width=\textwidth]{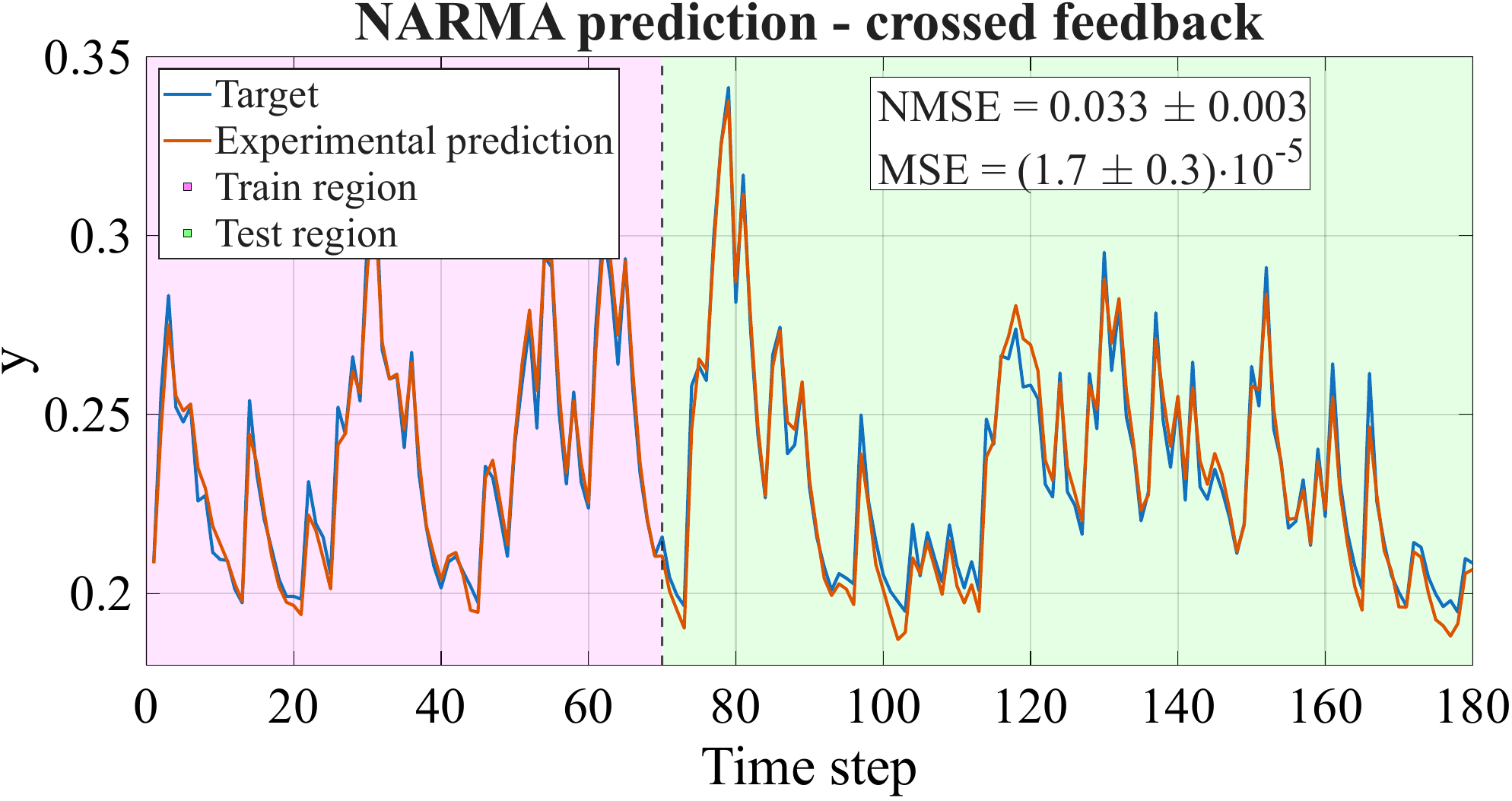}
        \subcaption*{(a)}
    \end{subfigure}
    \hfill
    \begin{subfigure}[b]{0.48\textwidth}
        \centering
        \includegraphics[width=\textwidth]{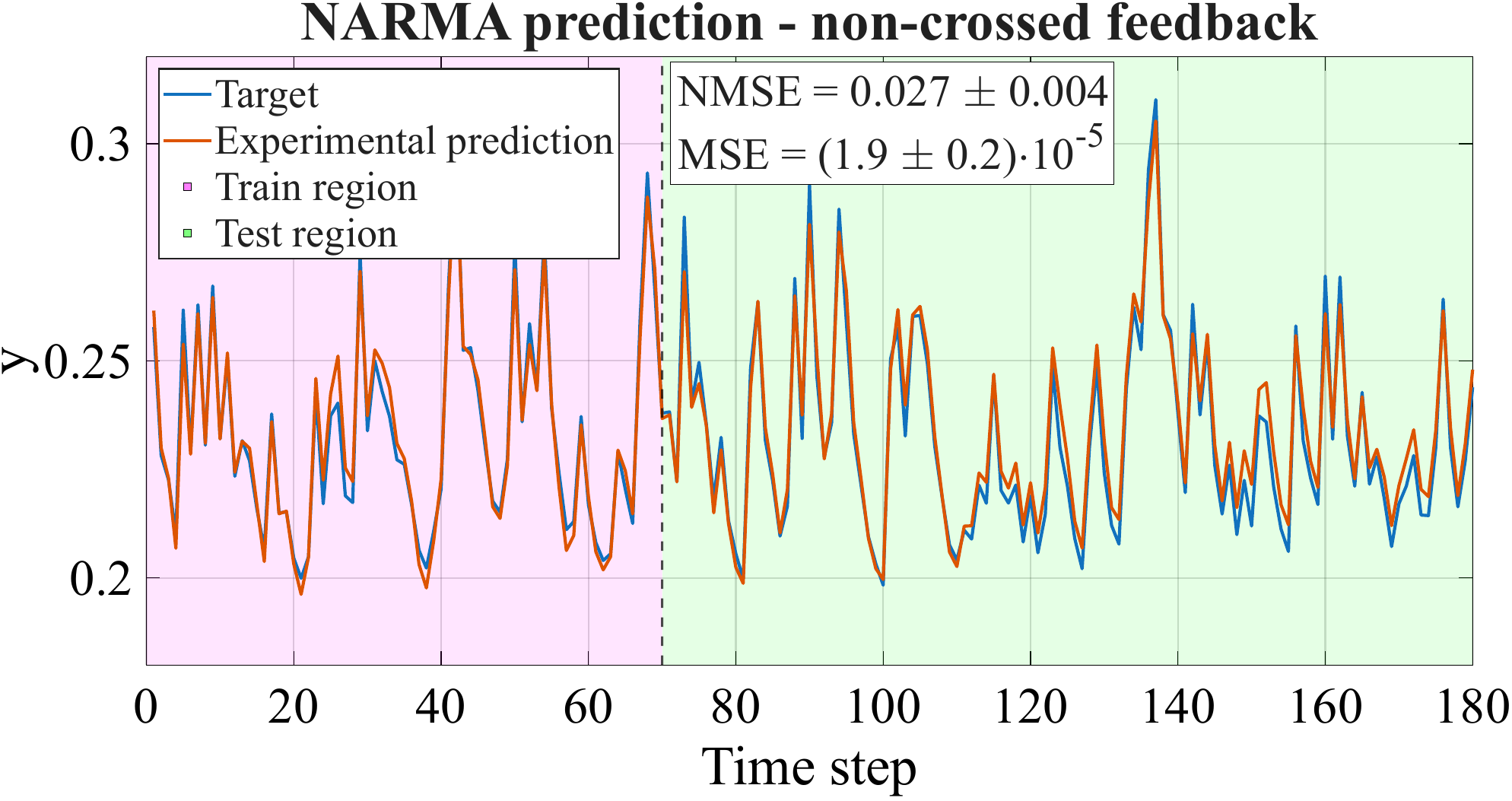}
        \subcaption*{(b)}
    \end{subfigure}    
    \caption{  Experimental prediction of NARMA time series using coupled memristors with (a) crossed and (b) non-crossed feedback laws. NARMA is defined as $y(t+1) = 0.4 y(t) + 0.4y(t)y(t-1) + 0.6x^3(t) + 0.1$, $x(t)$ being a uniformly distributed random number in the interval $[0,1/2]$~\cite{selimovic2025}. The modulation of $\langle N_{in}^{(1)}(t)\rangle$ (via MZI-1) is $x^2(t)$, while an additional $\pi/5$ phase shift is applied to MZI-2. The reflectivity update follows a moving-average-like feedback law: $R^{(i)}(t) = R^{(i)}(t-1) + [\langle N_{in}^{(j)}(t-1)\rangle - R^{(i)}(t-1)]/M^{(i)}$, where $i=[1,2]$ and $j$ can be equal (non-crossed feedback) or different (crossed feedback) to $i$. The acquired datasets are split into a training stage and a test stage. In the training stage, a linear regression of output photon fluxes ($X$) to $y$ is performed to obtain the coefficients $w$ such that $y = w\cdot X$. Then, the same coefficients $w$ are applied to the outputs in the test stage to predict the function. In the plots, the 'Experimental prediction' is $y_\text{pred}=w\cdot X$, which is an effective prediction only in the test stage. To assess the quality of the prediction, we estimate both the mean squared error (MSE = $\sum(y-y_{pred})^2/N$) and the normalized MSE (NMSE $=$ MSE$/\text{var}(y)$) over the test region. The two estimators are reported with their uncertainties obtained by statistics on 5 datasets.  The $\pi/5$ phase shift and the values of $M^{(1)}=3,M^{(2)} = 4$ for both feedback cases were chosen after performing simulations. These experiments have been performed with an attenuated CW laser with wavelength 750 nm.} 
    \label{fig:narma}
\end{figure}

Finally, our PQMs in parallel have been tested by using a common benchmark named nonlinear autoregressive moving average (NARMA)~\cite{ atiya2000new,wringe2025reservoircomputingbenchmarkstutorial}. This is a task to forecast the evolution of a time series which is commonly used to test memory and nonlinearities in various Machine Learning models, such as the Reservoir Computing~\cite{stepney2024physical}. 
We tested both configurations of the parallel PQMs, non-crossed and crossed feedback laws. Figure~\ref{fig:narma} reports the experimental results with the associated normalized mean squared error evaluated from five independent runs. Crossed (MSE=$(1.9\pm0.2)\cdot10^{-5}$) and non-crossed (MSE=$(1.7\pm0.3)\cdot10^{-5}$) feedback provide compatible performance, both providing an advantage with respect to the single memristor (MSE=$2.80\cdot 10^{-5}$ obtained in~\cite{selimovic2025}).  
An additional benchmark was tested, namely the NARMA-2 time series as defined in~\cite{di2026time}. In their work, a value of MSE$\,\simeq2.7\cdot10^{-3}$ is obtained by using two indistinguishable photons as the input of the neural network reservoir. To perform the same task, coupled PQMs provide MSE$\,=(2.2\pm0.2)\cdot10^{-3}$ and MSE$\,=(2.7\pm0.5)\cdot10^{-3}$, for crossed and non-crossed feedback, respectively.
Let us note that in this case the experiments have been carried out with an attenuated laser as a photon source. This was done in order to reduce the measurement time. In any case, as discussed in Section~\ref{sec:pic}, a weak classical light gives the same results of a single photon for the set of considered observables. These results and comparisons demonstrate the potential of two parallel PQMs to execute computational tasks that require nonlinearity and memory.

\section{Conclusions}
\label{sec:conclusions}

We have successfully implemented one and two coupled photonic quantum memristors using a small-scale silicon nitride photonic integrated circuit, combined with an off-chip silicon vacancy color center as a source of single photons, and external SPAD detectors. All components operate at room temperature and ambient pressure. Table~\ref{tab_comp} summarizes a comparison with other implementations of one or two photonic quantum memristors reported in the literature, showing that our work is the only one using a deterministic source and memristive elements both working at room temperature.  
Let us discuss the scaling opportunity for the proposed device. First, the insertion losses must be decreased: in our PIC, simple direct-tapering edge couplers and waveguides with 1 dB/cm propagation loss are utilized. However, in silicon nitride, more efficient solutions such as ultra-low-loss waveguides~\cite{Bauters_11}, low-loss inverted tapers for edge coupling \cite{SiNinvertedtapers} as well as high-efficiency grating couplers~\cite{di2026high} are nowadays commercially available at photonic foundries. These improvements allow to consider shorter acquisition time windows. Second, both the source and the single photon detectors~\cite{acerbi2024monolithically} should be on the same chip as the PIC to achieve a fully-integrated photonic quantum processor which can also be empowered by dedicated electronics to reconfigure the photonic devices and to manage the associated raw data.
While our nanodiamond-based single-photon source is currently implemented off-chip, deterministic positioning of nano-emitters has already been achieved in a broad range of nanophotonic and hybrid photonic platforms~\cite{acsami1c03905, adom202403288}. Furthermore, deterministic coupling of SiV$^-$ color centers~\cite{fehler2020purcell, Fehler2021HybridQuantumPhotonics}, together with their direct integration into silicon nitride photonic circuits~\cite{ngan2023quantum, Antoniuk2024AllOpticalSpinAccess}, has already been demonstrated.
While large scale PICs are already a reality, indistinguishable and pure single-photon sources based on colour centers are still under development and require low temperature~\cite{Martinez2022}.
Next, a still open challenge remains in the integration of multiphoton sources to increase the system's dimensionality.  

\begin{table}[!h]
    \centering
    \begin{tabular}{c||c c c}
           &~\cite{spagnolo2022experimental,selimovic2025} &~\cite{micco_quantum_2025}  & This work  \\ 
           \hline
          \hline
         \makecell{number of \\memristors} & 1 & 1 and 2 (cascaded) & 1 and 2 (in parallel) \\
         \hline
         \makecell{photon \\source} & \makecell{collinear Type II SPDC \\(heralding)}  & quantum dot~\cite{Somaschi_2016} &  SiV$^-$ color center  \\
         \hline
         \makecell{memristor \\implementation} & \makecell{FLW-waveguides \\in silicate glass~\cite{Ceccarelli_2020}}  &  bulk optical setup  & Silicon nitride PIC \\
         \hline
         \makecell{working \\temperature} & $\sim$300 K & $\sim$4 K (the source) & $\sim$300 K
    \end{tabular}
    \caption{Comparison of photonic quantum memristor experiments in terms of the number of devices, the type of source, the photonic platform used for the implementation, and working temperature.
    FLW stands for femtosecond-laser written, and SPDC for spontaneous parametric down conversion. }
    \label{tab_comp}
\end{table}

Our single PQM measurement results reproduce very closely those reported in~\cite{spagnolo2022experimental}, where a bulk crystal was instead employed to generate single photons via spontaneous parametric downconversion followed by heralding. Moreover, we present experiments of two coupled PQMs featuring a cross-feedback law. The results show different hysteresis curves, depending on the fact that intra- or inter-memristor relations are considered, characterized by novel features, among them: non-pinched behaviour, enhanced form factor and self-intersecting hysteresis geometry. Our experiments clearly highlight an interesting response of these devices as building blocks for quantum neural networks. Indeed, the correlations introduced in the crossed feedback laws, Eq.~\eqref{eq:nest_feed_law}, can enhance the nonlinearity and the history-dependent response of the overall structure. The presented simulations make manifest the richness of the new dynamics, arising from the classical correlations between the memristors and their inputs.

These results mark a step forward towards integrated photonic quantum reservoir computing~\cite{sakurai2025quantum} built by means of universal MZI schemes~\cite{reck_experimental_1994,clements_optimal_2016} and photonic quantum memristors~\cite{spagnolo2022experimental}. 
By showing that crossed feedback laws between two memristors modify both intra- and inter-memristor input-output relations, our work demonstrates that non-trivial network dynamics can be engineered even in a minimal coupled architecture. This suggests a route towards compact photonic reservoirs in which dynamical richness is enhanced not only by increasing the number of physical nodes, but also by tailoring the feedback connectivity between memristive elements. Moreover, we highlight that quantum correlations within the quantum reservoir hidden layers made of PQMs can lead to complex high-dimensional configuration spaces without the need for large and energy-consuming physical systems as reservoirs.

At the same time, reservoir complexity must be matched to the target task: excessively complex reservoirs may improve the separation of training histories while reducing generalization performance, leading to overfitting~\cite{dutoit2009pruning,gauthier2021next,ito2025overfitting}. Strategies such as readout regularization and node pruning will therefore be relevant when scaling these architectures. As shown in recently reported  experiments~\cite{spagnolo2022experimental,micco_quantum_2025,selimovic2025,di2026time} and in the NARMA benchmark reported in Section~\ref{sec:results}, adaptive feedback provides effective nonlinear and history-dependent dynamics that are well suited to time-series processing. Since only the classical readout layer is trained, this approach fits naturally within NISQ reservoir computing platforms ~\cite{mujal2021opportunities,abbas2024classical,selimovic2025,di2026time}. In contrast to variational quantum algorithms~\cite{Peruzzo_2014,McClean_2016, cerezo,Tilly_2022}, it avoids the direct optimization of large sets of quantum parameters associated with the preparation of trial states, thereby avoiding a major source of trainability issues such as expressivity-induced barren plateaus~\cite{Tangpanitanon_2020,Holmes_2022,mcclean2018barren}.

\newpage

\section*{Acknowledgements}
We are grateful for financial supports from Horizon Widera 2023 (101160101) through ToEQPL project. R.P.G.K.A. and C.C. are grateful for financial supports from the EUR NANO-PHOT (ANR 18-EURE-0013) as well as the Region Grand Est and the OQuLus PEPR project (ANR 22-PETQ-0013).

\section*{Author contributions}
A.B. and R.P.G.K.A. contributed equally to this work.
R.P.G.K.A.,  and R.D. performed the optical characterization of the single photon source at UTT using a $\mu$-PL setup. R.P.G.K.A. and Z.A.A. assembled a similar more portable $\mu$-PL setup at University of Trento and used the home-made triangulation protocol to find the same single photon emitter. 
A.B., R.P.G.K.A., S.G., X.C., Z.A.A. carried out the memristors' experiments. 
I.F. performed the time series prediction experiments.
A.B. and S.G. analysed the results of the memristors' experiments and performed the simulations.
S.A., C.C. and L.P. provided critical feedback and helped shape the research, analysis and manuscript.

\section*{Competing interests}
The authors declare no competing financial interests.



\clearpage

\appendix

\section*{Appendix}

\section{Setup}
\label{app:setup}
\begin{figure}[ht]
  \centering
  \includegraphics[width=\linewidth]{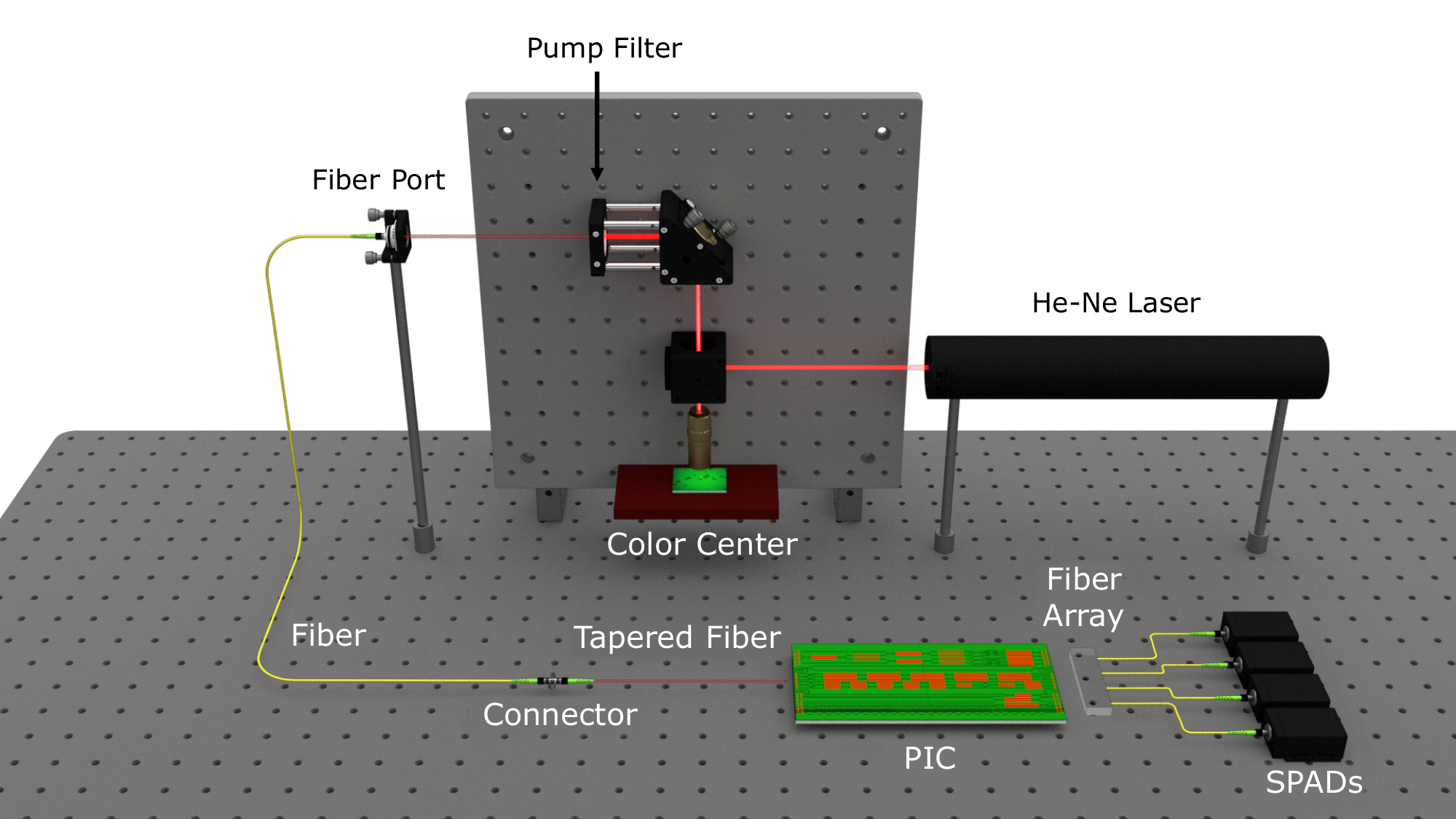}
  \caption{A simplified image of the setup.}
  \label{fig:setup}
\end{figure}
The PL and correlation measurements were made at UTT using a 633 nm Helium Neon laser to excite the SiV$^-$ through a 100X air objective with numerical aperture of 0.95 for optimal collection of the emitted photons. The laser is first directed to a pellicle beam splitter with a reflection/transmission ratio of 8:92. Part of the laser is reflected to the objective, as shown in Fig.~\ref{fig:setup}. The focused laser beam excites the SiV$^-$ and the fluorescence is collected by the same objective and sent to the collection path via a flipping mirror that allowed to choose between a spectrometer or a HBT setup.  A laser cleanup filter (Semrock LL01-633-25) was placed along the illumination path to the sample while a long-pass filter at 647 nm (Semrock LP02-647RU-25) was used along the collection path to reject the laser. An additional narrowband filter at 740 nm (Semrock FF01-740/13-25) was also used for the HBT experiments. The setup assembled at the University of Trento for the experiments reported in this work (see Fig.~\ref{fig:setup}) follows a similar design, with the only change being the laser source, which was a 640 nm diode laser. A clean-up filter (Semrock VersaChrome 704/13 tunable bandpass filter) was used immediately after the laser and the same filter combination as on the collection path of the setup at the UTT was used at the end of the collection arm to reject the laser. A fiber port was used to couple the collected emission to a single mode fiber that fed the PIC. Optical coupling at the chip's input is performed through a tapered lensed fiber, while output coupling is realized using a standard fiber array. The photonic circuit was made via a photolithographic process through a commercial service provided by Ligentec SA. The waveguides are made of a silicon nitride core, 150 nm thick and 550 nm wide, embedded in a silica cladding. This cross-section ensures single-mode propagation at wavelengths around $750$ nm, and thus at the emission wavelength of the source. The total insertion loss of the photonic circuit is around 20 dB. A Peltier Cell is utilized to stabilize the PIC temperature, and it is managed by a PID controller. An electrical power supply (Qontrol systems - BP8 device model) provides the currents for all the thermal phase shifters (PSs) of the PIC. Single photons are detected by silicon SPADs (Excelitas), all featuring the same efficiencies and dark counts. Finally, the output counts are managed by a fast time-tagging electronics (Swabian Instruments) connected to a PC.
The total input photon flux is around $10^2$ kHz, while the output flux is around 1 kHz. We chose a time bin of $2$ s and $20$ s, denoted as $\tau$, during the calibration routine and the memristor experiments, respectively.

\section{Triangulation}
\label{app:Triangulation}

This a technique we developed to reliably locate previously identified single photon emitters on the substrates. It is based on an inspection camera and on nano-positioner stages. Our protocol consists of two main parts: acquiring the reference frame information and then imposing a transformation on it. In the initial frame of reference, the $(x,y)$ coordinate readout from the nanopositioner stages are recorded for at least 3 reference points $(x_1,y_1)$, $(x_2,y_2)$ and $(x_3,y_3)$, which are usually chosen to be any clear and sharp feature near the corners of the substrate, and for the  point of interest $(x_p,y_p)$.

A linear transformation corresponds to the relocation of the sample with the emitter in a new setup. First, we make sure that the same reference points can be retrieved after the linear transformation  by recording images of each reference corner with the exact sharp feature made to coincide with the center of the camera crosshair overlay. Then, we readout the new set of coordinates $(x'_1,y'_1)$, $(x'_2,y'_2)$ and $(x'_3,y'_3)$ for each of the reference points. These are used to get the coordinates of the point of interest in the new transformed reference frame. 

\begin{figure}[ht]
  \centering
  \includegraphics[width=\linewidth]{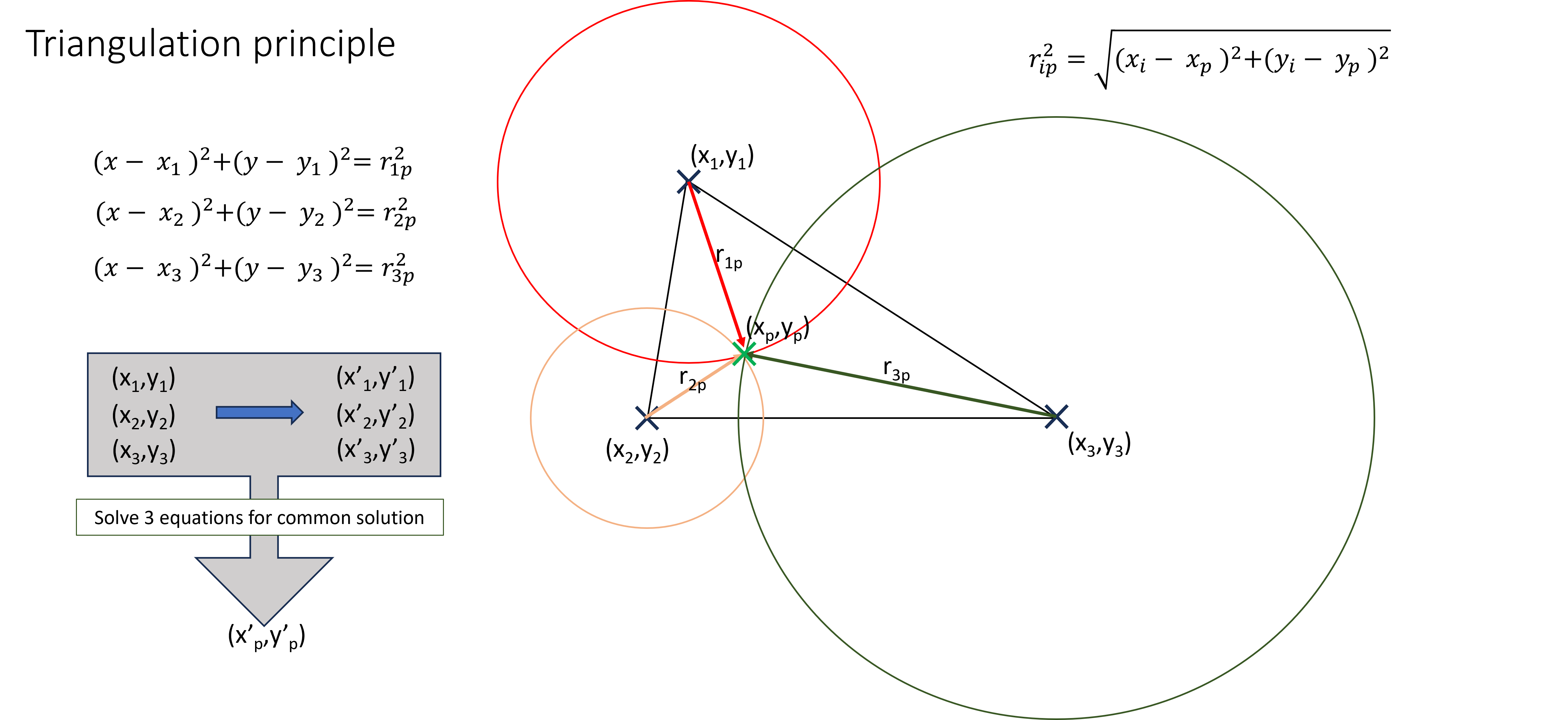}
  \caption{The basic principle behind triangulation.}
  \label{fig:Triangulation}
\end{figure}

The reference points in the old frame can be thought of as the centers of three circles whose radii equal their respective distances from the point of interest $(x_p,y_p)$. The Euclidean distance can easily be calculated for each of the three distances $r_{1p}$, $r_{2p}$ and $r_{3p}$, as shown in Fig.~\ref{fig:Triangulation}. With these radii,  the equations of the three circles can be defined, as shown in Fig.~\ref{fig:Triangulation}. The relation between these four points remains the same even after the linear transformation, which means that we now have the same set of three circles in the new frame, with different centers but the same radii. Solving these three equations for a common solution gives the coordinates for the point of interest $(x'_p,y'_p)$ in the new frame of reference, since three non-concentric circles can intersect at only one point. Finally, the nano-positioner stages move to this point and the emitter should be within the field of view of the inspection camera. 

There are limitations to this technique, particularly when it comes to emitters with dimensions smaller than the diffraction limit.  While the reference coordinates corresponding to the corners can still easily be acquired, it is difficult to zero in on the nano-emitter which cannot be resolved on the inspection camera. Another problem is the accuracy, repeatability and positional instability or drift of the nanopositioner stages. A quick work around for these problems is to use triangulation to arrive in the neighborhood of the emitter and then proceed with a confocal scan to find it.

\section{Procedures for memristors experiments}
\label{app:coupled_mems}

\begin{figure}[ht]
  \centering
  \includegraphics[width=\linewidth,trim=0cm 0.1cm 0cm 0cm,clip]{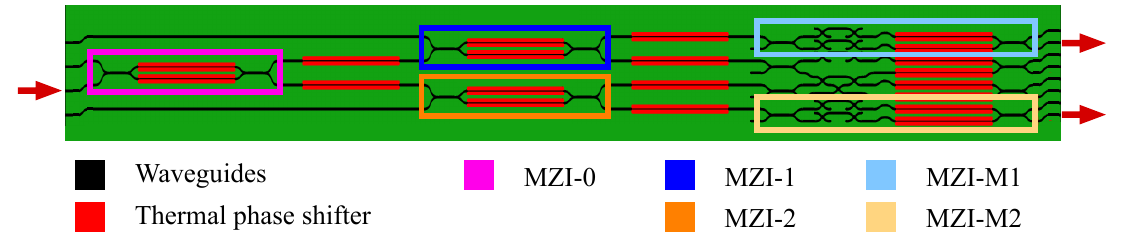}
  \caption{The layout of the circuit implementing single memristor and two coupled memristors. Only the highlighted Mach Zehnder interferometers (MZIs) are utilized. The arrow on the left indicates the input waveguide (third one from the top), while the arrows on the right are placed to highlight the output waveguides (the second and eighth ones from the top), fiber-coupled to single-photon avalanche diodes.}
  \label{fig:pic_scheme_app}
\end{figure}

Fig.~\ref{fig:pic_scheme_app} shows the schematic layout of the used PIC, which features four input and eight output waveguides. 
The circuit is composed of two sections: a triangular scheme of three MZIs, denoted as MZI-0, MZI-1 and MZI-2, and a network composed of MMIs, crossing waveguides and thermal phase shifters, containing MZIs denoted as MZI-M1 and MZI-M2. This circuit was originally designed to implement a swap test on a generic state of two qubits based on single-photon path encoding~\cite{baldazzi2024linear}. In Appendix~E of~\cite{baldazzi2024linear}, the characterization curves of MMIs are reported. \\
We utilize only the MZIs highlighted in Fig.~\ref{fig:pic_scheme_app}. We choose to inject the single photons into the third input waveguide (counting from the top of the layout), and collect them with a fiber array. Only the second and eighth output waveguides (counting from the top) are connected to fiber-coupled single-photon silicon avalanche diode (SPAD) detectors.

In the single-memristor experiments, the following steps are performed within each time bin $t_k = k \tau $ with $k\in [0,N-1]$ (refer to Fig.~\ref{fig:pic_scheme_app}):
\begin{enumerate}
\item Set the MZI-0 to modulate the input flux with transmission equal to $\sin^2\left(\pi t_k/ T_{\rm osc}\right)$, and set MZI-1 and MZI-2 to transmit all the photons to the upper and lower output, respectively.
\item Set the MZI-M1 implementing the memristor with phase $\theta_k$, which corresponds to transmission to its lower output waveguide equal to $R(t_k)$, and set MZI-M2 to transmit all the photons to the eighth output waveguide.
\item Collect single-photon detection events at the detectors fiber-coupled to the two output waveguides for $\tau_{meas}<\tau$.
\item Set the MZI-M1 implementing the memristor with phase $\theta_k+\pi/2$, corresponding to transmission to the lower output waveguide equal to $1-R(t_k)$.
\item Collect single-photon detection events at the detectors fiber-coupled to the two output waveguides for $\tau_{meas}<\tau$.
\item Evaluate $\langle N_{in}(t_k)\rangle$, $\langle N_{out}(t_k)\rangle$, and update the phase setting of the memristor MZI-M1 according to the chosen feedback law.
\item Return to step 1.
\end{enumerate}
In the first time bin, the parameter $R$ of the memristor is set to zero, and the memristor buffer is composed of a null vector of dimension $M$. After $M$ time bins, the buffer is completely filled, and new entries are saved by keeping the last $M$ entries (LIFO). After the buffer is full, the memristor starts to show hysteresis. This procedure could be implemented with just one detector by exploiting the $\pi/2$-relation of the MZI modulating the input. However, in order to reduce the duration of the measurement, for the single-memristor case two output waveguides are measured. Instead, for the coupled memristors case, only one output channel per memristor is used.
The input and output averaged photon fluxes are 
\begin{equation}
\begin{split}
& \langle N_{in}(t_k)\rangle = \frac{ N_2(\theta_k) + N_2(\theta_k+\pi/2) }{ N_{tot}(t_k) } \quad,\quad
\langle N_{out}(t_k)\rangle = \frac{ N_2(\theta_k+\pi/2) }{ N_{tot}(t_k) } \,, \\
& \quad \mbox{where} \,\,\, N_{tot}(t_k) = N_2(\theta_k) +  N_2(\theta_k+\pi/2) +  N_8(t_k) \,,
\end{split}
\end{equation}
where $N_2(\theta_k)$ and $N_2(\theta_k+\pi/2)$ are the number of detector's clicks at the second output during the step 3 and 5 of the $k$-th time bin, respectively, while $N_8(t_k)$ the number of detector's clicks at the eighth output during step 3.
The feedback law for MZI-M1 is 
\begin{equation}
\begin{split}
& R(t_k) = \frac{1}{2} + \frac{1}{M} \sum_{j=k-M+1}^k \left( \langle N_{in}(t_j) \rangle -\frac{1}{2} \right)\,,
\end{split}
\end{equation}
which is the same used in the first implementation of a photonic quantum memristor~\cite{spagnolo2022experimental}.

\begin{figure}[ht]
  \centering
  \includegraphics[width=0.4\linewidth]{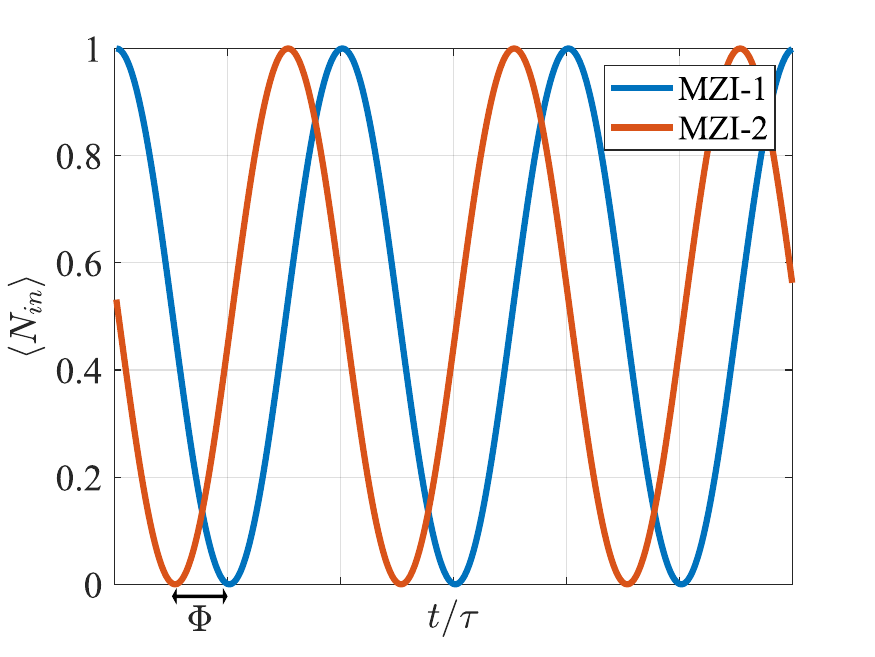}
  \caption{The dependence of the input fluxes for the two parallel memristors. MZI-1 and MZI-2 are the MZI used to modulate the input fluxes of photons, Fig.~\ref{fig:pic_scheme_app}, and they are shifted by a relative phase $\Phi$.}
  \label{fig:input_mod}
\end{figure}

In the coupled-memristor experiments, the following steps are performed within each time bin $t_k = k \tau $ with $k\in [0,N-1]$ (refer always to Fig.~\ref{fig:pic_scheme_app}):
\begin{enumerate}
\item Set MZI-0 to equally split the input flow to its two outputs, and set MZI-1 and MZI-2 to modulate the input fluxes with transmission equal to $\sin^2\left(\pi t_k/ T_{\rm osc}\right)$ and $\sin^2\left(\pi t_k/ T_{\rm osc}+\Phi\right)$, respectively, where $\Phi$ is a phase difference between the two inputs, Fig.~\ref{fig:input_mod}.
\item Set MZI-M1 and MZI-M2 with phases $\theta_k^{(1)}$ and $\theta_k^{(2)}$, which correspond to transmission to their lower output waveguides equal to $R^{(1)}(t_k)$ and $R^{(2)}(t_k)$, respectively.
\item Collect single-photon detection events for $\tau_{meas}<\tau$ from the two output waveguides.
\item Set MZI-M1 and MZI-M2 with phases $\theta_k^{(1)}+\pi/2$ and $\theta_k^{(2)}+\pi/2$, which correspond to transmission to their lower output waveguides equal to $1-R^{(1)}(t_k)$ and $1-R^{(2)}(t_k)$, respectively.
\item Collect single-photon detection events for $\tau_{meas}<\tau$ from the two output waveguides.
\item Set MZI-1 and MZI-2 with transmission equal to $\cos^2\left(\pi t_k/ T_{\rm osc}\right)$ and $\cos^2\left(\pi t_k/ T_{\rm osc}+\Phi\right)$ respectively (both have $\pi/2$-shift with respect to step 1), and set MZI-M1 and MZI-M2 to transmit all the photons to the lower output waveguides, outputs two and eight respectively.
\item Collect single-photon detection events for $\tau_{meas}<\tau$ from the two output waveguides.
\item Evaluate $\langle N_{in}^{(1)}(t_k)\rangle$, $\langle N_{out}^{(1)}(t_k)\rangle$, $\langle N_{in}^{(2)}(t_k)\rangle$, $\langle N_{out}^{(2)}(t_k)\rangle$ and update the settings of the memristors according to the chosen feedback law.
\item Return to step 1.
\end{enumerate}
The described procedure uses the minimum number of detectors, one per memristor. Also in this case, the parameter $R$ of the memristor is initially set to zero, the memristors buffer is made of a null vector of dimension $M$, and new entries in the buffers are saved by keeping the last $M$ entries (LIFO). The input and output averaged photon fluxes are 
\begin{equation}
\begin{split}
& \langle N_{in}^{(1)}(t_k)\rangle = \frac{ N_2(\theta_k^{(1)}) + N_2(\theta_k^{(1)}+\pi/2) }{ N^{(1)}_{tot}(t_k) } \quad,\quad
\langle N_{out}^{(1)}(t_k)\rangle = \frac{ N_2(\theta_k^{(1)}+\pi/2) }{ N^{(1)}_{tot}(t_k) } \,, \\
& \langle N_{in}^{(2)}(t_k)\rangle = \frac{ N_8(\theta_k^{(2)}) + N_8(\theta_k^{(2)}+\pi/2) }{ N^{(2)}_{tot}(t_k) } \quad,\quad
\langle N_{out}^{(2)}(t_k)\rangle = \frac{ N_8(\theta_k^{(2)}+\pi/2) }{ N^{(2)}_{tot}(t_k) } \,, \\
& \quad \mbox{where} \,\,\, N^{(1)}_{tot}(t_k) = N_2(\theta_k^{(1)}) +  N_2(\theta_k^{(1)}+\pi/2) +  \bar{N}_2(t_k) \,, \\
& \quad \quad \mbox{and} \,\,\, N^{(2)}_{tot}(t_k) = N_8(\theta_k^{(2)}) +  N_8(\theta_k^{(2)}+\pi/2) +  \bar{N}_8(t_k) \,, \\
\end{split}
\end{equation}
where $N_{2/8}(\theta_k^{(1/2)})$ and $N_{2/8}(\theta_k^{(1/2)}+\pi/2)$ are the number of detector's clicks at the second/eighth output during the step 3 and 5 of the $k$-th time bin, respectively, while $\bar{N}_{2/8}(t_k)$ the number of detector's clicks at the second/eighth output during step 7.

\section{Simulations of the experiments}
\label{app:simulations}

The memristor consists of a MZI. Its reflectivity ideally should follow the law~\cite{spagnolo2022experimental}:
\begin{equation}
\begin{split}
& R_{mem}(t_k) = \frac{1}{2} + \frac{1}{M} \sum_{j=k-M+1}^k \left( \langle N_{in} (t_j)\rangle -\frac{1}{2} \right) \,.
\end{split}
\end{equation}
Practically, $R$ is set by imposing a relative phase $\phi$ between the two arms of the MZI:
\begin{equation}
\begin{split}
& R = \sin^2\left( \frac{\phi}{2} \right) \,.
\label{eq:Rappsimu}
\end{split}
\end{equation}
The ideal average values $\langle N_{in} \rangle$ and $\langle N_{out} \rangle$ read as follows
\begin{equation}
\begin{split}
& \langle N_{in}(t_k) \rangle = \sin^2\left( \frac{\phi^{in}(t_k)}{2} \right) \,, \\
& \langle N_{out}(t_k) \rangle = \left[ 1- R_{mem}(t_k)\right] \langle N_{in}(t_k) \rangle \,.
\end{split}
\end{equation}

From an experimental point of view, there could be some error in the setting of
$\phi$ for the MZIs. In particular, we can model the real set phase as
\begin{equation}
\begin{split}
& \phi_{\rm real} = \phi_{\rm theo} + \delta\phi_{\rm sta} + \delta\phi_{\rm err} \,,
\end{split}
\end{equation}
where $\phi_{\rm theo} = 2 \,\mbox{arcsin}\sqrt{R}$ is the ideal value of $\phi$, while $\delta\phi_{\rm sta}$ and $\delta\phi_{\rm err}$ represent a systematic static offset (due to calibration errors) and stochastic noise (due to temperature
or current fluctuations), respectively.
Another source of non-ideality is the polarization of the photons entering the
chip. The integrated components are designed to work with TE light. Polarization
components different from TE lead to non-ideal behaviours. In particular,
they eventually affect the visibilities of the MZIs. Therefore, in practice we can model $R$ as an effective reflectivity given by
\begin{equation}
\begin{split}
& R = \frac{1}{2} \left[ 1- V \cos\left( \phi_{\rm real} \right) \right] \,,
\label{eq:Rnonidappsimu}
\end{split}
\end{equation}
where $V$ is the visibility of the interference fringe associated to one MZI. $V$ can be extracted from the calibration data together with a confidence interval.
Finally, we have to deal with non-idealities due to dark counts, noise and detector efficiency. 

Therefore, we can model the non-idealities as follows
\begin{equation}
\begin{split}
& \langle N_{in}(t_k) \rangle = \frac{1}{2} \left[ 1- V_{in} \cos\left( \phi^{in}_{\rm real}(t_k) \right) \right] \,, \\
& \langle N_{out} (t_k)\rangle = \eta \left[ 1- R_{mem}(t_k)\right] \langle N_{in}(t_k) \rangle + N_{\rm dark} + \xi_{\rm err} (t_k)\,,
\label{eq:appsimNresults}
\end{split}
\end{equation}
where $R_{mem}$ is given by Eq.~\eqref{eq:Rnonidappsimu}, $\eta$ is the detector efficiency, $N_{\rm dark}$ represents dark counts and $\xi_{\rm err}$ is stochastic noise.

The simulated results shown in Fig.~\ref{fig:single_mem} and Fig.~\ref{fig:coupled_mems} are obtained by putting to zero the stochastic noises and using the parameters reported in Table~\ref{tab:single_mem} and Table~\ref{tab:coupled_mems}, respectively. These parameters are found through an optimization routine by keeping the visibilities bounded by the value and associated error founded in the calibration routine and the static systematic phase error of the order of $0.1$ rad.
Each row of the two tables corresponds to a different experiment and a different calibration of the thermal phase shifters.\\
The optimal values for the experimental parameters reported in Table~\ref{tab:single_mem} and Table~\ref{tab:coupled_mems} are different depending on the devices configuration, since the polarization changes. The change is due to the fact that between every experiment, the source rate and the PIC-fiber coupling are optimized, and this procedure implies the movement of the fiber connecting the source and the PIC. Since there is no polarization controller due to the high PIC insertion losses and the fibers are not polarization maintaining, it is not possible to set the same polarization in the different experiments. Finally, regarding the polarization stability of the experiments, room temperature changes affects the input polarization. However, this effect can be well-described by the stochastic noise terms in Eq.~\eqref{eq:appsimNresults}, whose effect falls within the experimental error. \\
We conclude by saying that the optimal parameters in Table~\ref{tab:single_mem} and Table~\ref{tab:coupled_mems} can be understood either as mean-temporal values or values different from the real experimental ones but able to fit the experimental outcomes. This second possibility cannot be excluded, since the optimization routine is performed over a fuzzy configuration landscape and the stochastic noise has been integrated out from the optimization. 

\begin{table}[ht]
  \centering
  \caption{Parameters used for the simulated results of a single memristor shown in Fig.~\ref{fig:single_mem}. The superscripts $(0)$ and $({\rm M1})$ stand for MZI-0 and MZI-M1, respectively. The static systematic phase errors $\delta\phi_{\rm sta}$ are measured in radians. The parameters are rounded to the second decimal place.}
  \label{tab:single_mem}
  \begin{tabular}{lcccc}
    \toprule
     & $V^{(0)}$ & $V^{\rm (M1)}$ & $\delta\phi_{\rm sta}^{\rm (0)}$ & $\delta\phi_{\rm sta}^{\rm (M1)}$\\
    \midrule
    $T=0.01\,T_{\rm osc}$ & 0.99 & 0.83 & -0.03 & -0.11 \\
    $T=0.1\,T_{\rm osc}$  & 1.00 & 0.82 & 0.05 & -0.20 \\
    $T=0.3\,T_{\rm osc}$  & 1.00 & 0.92 & 0.03 & 0.09 \\
    $T=0.5\,T_{\rm osc}$  & 0.95 & 0.92 & 0.00 & 0.03 \\
    $T=0.7\,T_{\rm osc}$  & 0.93 & 0.80 & 0.17 & -0.16 \\
    $T=T_{\rm osc}$       & 0.99 & 0.90 & 0.12 & -0.20 \\
    \bottomrule
  \end{tabular}
\end{table}

\begin{table}[ht]
  \centering
  \caption{Parameters used for the simulated results of the coupled memristors shown in Fig.~\ref{fig:coupled_mems}. The superscripts $(0)$, $(1)$, $(2)$, $({\rm M1})$ and $({\rm M2})$ stand for MZI-0, MZI-1, MZI-2, MZI-M1 and MZI-M2, respectively. The static systematic phase errors $\delta\phi_{\rm sta}$ are measured in radians. The parameters are rounded to the second decimal place.}
  \label{tab:coupled_mems}
  \begin{tabular}{lccccccccc}
    \toprule
     & $V^{(0)}$ & $V^{(1)}$ & $V^{(2)}$ & $V^{\rm (M1)}$ & $V^{\rm (M2)}$ & $\delta\phi_{\rm sta}^{(1)}$ & $\delta\phi_{\rm sta}^{(2)}$ & $\delta\phi_{\rm sta}^{\rm (M1)}$ & $\delta\phi_{\rm sta}^{\rm (M2)}$  \\
    \midrule
    $T=0.2\,T_{\rm osc}$ and $\Phi=0.7$rad  & 0.95 & 0.83 & 0.89 & 0.95 & 0.91 & -0.16 & -0.02 & 0.03 & -0.02 \\
    $T=0.3\,T_{\rm osc}$ and $\Phi=0.5$rad  & 1.00 & 0.90 & 0.95 & 1.00 & 0.87 & -0.10 & -0.19 & -0.01 & -0.02 \\
    $T=0.4\,T_{\rm osc}$ and $\Phi=0.7$rad  & 0.95 & 0.92 & 0.97 & 1.00 & 0.89 & -0.20 & -0.02 & 0.01 & 0.00 \\
    \bottomrule
  \end{tabular}
\end{table}

\section{Coherent parallel memristors}
\label{app:quantumcoupledPQMs}

Let's consider the case of a single photon state passing through a 50-50 beam splitter, whose output channels are connected to the inputs of two different reconfigurable beam splitters, each able to prepare the state in Eq.~\eqref{eq:initialstate_singlePQM}. 
Looking at Fig.~\ref{fig:pic_scheme} and Fig.~\ref{fig:pic_scheme_app}, in the coupled-memristor experiment, the 50-50 beam splitter is MZI-0 and the two reconfigurable beam-splitter are MZI-1 and MZI-2.
We denote by A and B the output channels of MZI-1 and by D and E the output channels of MZI-2.
At this stage, the state reads as follows
\begin{equation}
    |\psi_{in}\rangle = \frac{1}{\sqrt{2}} \left[ |\psi_{in}^{(1)}\rangle\otimes|0^{(2)}\rangle +
    {\rm e}^{{\rm i} \delta}|0^{(1)}\rangle\otimes|\psi_{in}^{(2)}\rangle\right] \,,
\end{equation}
where $\delta$ is a relative phase at the output of the balanced beam splitter and
\begin{equation}
\begin{split}
    |\psi_{in}^{(1)}\rangle = \alpha_1(t) |1_{\rm A},0_{\rm B}\rangle + \beta_1(t) |0_{\rm A},1_{\rm B}\rangle \quad,\quad |0^{(1)}\rangle = |0_{\rm A},0_{\rm B}\rangle \\
    |\psi_{in}^{(2)}\rangle = \alpha_2(t) |1_{\rm D},0_{\rm E}\rangle + \beta_2(t) |0_{\rm D},1_{\rm E}\rangle \quad,\quad |0^{(2)}\rangle = |0_{\rm D},0_{\rm E}\rangle
    \end{split}
\end{equation}
and the qubit state $|0_{\rm A/B/D/E}\rangle$ corresponds to zero photons in mode A/B/D/E and $|1_{\rm A/B/D/E}\rangle$ corresponds to a photon in mode A/B/D/E. As usual, $|\alpha_{1/2}(t)|^2+|\beta_{1/2}(t)|^2=1$.
The input density matrix is simply $\rho_{in}(t)\equiv|\psi_{in}(t)\rangle\langle\psi_{in}(t)|$.

Then, two auxiliary modes denoted by C and F without photons are added to the input state, giving the total state $|\psi_{in}(t)\rangle\otimes|0_{\rm C}\rangle\otimes|0_{\rm F}\rangle$. Then, we apply the transformations $U_{1/2}(t)$ associated with the MZI-M1 and MZI-M2, i.e. the two memristors in parallel, acting on the mode pairs (B,C) and (E,F), respectively. The result reads $U_1(t)\otimes U_2(t) \,\rho_{in}(t) \otimes|0_{\rm C}\rangle\langle0_{\rm C}|\otimes|0_{\rm F}\rangle\langle0_{\rm F}|\, \left(U_1(t)\otimes U_2(t)\right)^\dagger$.
Given the input state and denoting the reflectance of the MZI-M1/2 as $R_{1/2}(t)$, the MZI-M1/2 map reads: 
\begin{equation}
\begin{split}
     U_{1/2}(t) &= \mathbf{1}_{\rm A/D}\otimes \left[|0_{\rm B/E},0_{\rm C/F}\rangle\langle 0_{\rm B/E},0_{\rm C/F}|
     \right. \\
     & \hspace{2cm}\left.+\sqrt{1-R_{1/2}(t)}\,|1_{\rm B/E},0_{\rm C/F}\rangle\langle 1_{\rm B/E},0_{\rm C/F}|+\sqrt{R_{1/2}(t)}\,|0_{\rm B/E},1_{\rm C/F}\rangle\langle 1_{\rm B/E},0_{\rm C/F}| \right. \\
     & \hspace{2cm}\left. +\sqrt{1-R_{1/2}(t)}\,|0_{\rm B/E},1_{\rm C/F}\rangle\langle 0_{\rm B/E},1_{\rm C/F}|-\sqrt{R_{1/2}(t)}\,|1_{\rm B/E},0_{\rm C/F}\rangle\langle 0_{\rm B/E},1_{\rm C/F}| \right]\,, 
\end{split}
\end{equation}
where $\mathbf{1}_{\rm A/D}$ is the identity map over the mode A/D. We use the same convention of the paper: a ``transmitted photon'' is a photon entering in mode B/C or E/F and remaining in mode B/C or E/F after the MZI with probability $1-R_{1/2}(t)$, while a ``reflected photon'' is a photon entering in mode B/C or E/F and exiting in mode C/B or F/E with probability $R_{1/2}(t)$.
The output density matrix $\rho_{out}(t)$ is defined as the partial trace over the modes C and F of the density matrix $U_1(t)\otimes U_2(t) \,\rho_{in}(t) \otimes|0_{\rm C}\rangle\langle0_{\rm C}|\otimes|0_{\rm F}\rangle\langle0_{\rm F}|\, \left(U_1(t)\otimes U_2(t)\right)^\dagger$. This means that 
\begin{equation}
\begin{split}
    \rho_{out}(t) &\equiv \langle0_{\rm C},0_{\rm F} |U_1(t)\otimes U_2(t) \,\rho_{in}(t) \otimes|0_{\rm C}\rangle\langle0_{\rm C}|\otimes|0_{\rm F}\rangle\langle0_{\rm F}|\, \left(U_1(t)\otimes U_2(t)\right)^\dagger|0_{\rm C},0_{\rm F}\rangle \\
    & \hspace{1cm}+\langle 1_{\rm C},0_{\rm F} |U_1(t)\otimes U_2(t) \,\rho_{in}(t) \otimes|0_{\rm C}\rangle\langle0_{\rm C}|\otimes|0_{\rm F}\rangle\langle0_{\rm F}|\, \left(U_1(t)\otimes U_2(t)\right)^\dagger|1_{\rm C},0_{\rm F}\rangle \\
    & \hspace{1cm}+\langle 0_{\rm C}, 1_{\rm F} |U_1(t)\otimes U_2(t) \,\rho_{in}(t) \otimes|0_{\rm C}\rangle\langle0_{\rm C}|\otimes|0_{\rm F}\rangle\langle0_{\rm F}|\, \left(U_1(t)\otimes U_2(t)\right)^\dagger|0_{\rm C}, 1_{\rm F}\rangle 
    \\
    & = \frac{1}{2}\left[ \rho_{out}^{(1)}(t) \otimes |0_{\rm D},0_{\rm E}\rangle\langle0_{\rm D},0_{\rm E}| 
    + |0_{\rm A},0_{\rm B}\rangle\langle0_{\rm A},0_{\rm B}| \otimes \rho_{out}^{(2)}(t) + {\rm e}^{-{\rm i} \delta}\rho^{(mix)}+ {\rm e}^{{\rm i} \delta}\left(\rho^{(mix)}\right)^\dagger
    \right]\,.
\end{split}
\end{equation}
where
\begin{equation}
\begin{split}
    \rho_{out}^{(1)}(t) 
    & = \left( \alpha_{1}(t)|1_{\rm A},0_{\rm B}\rangle +\beta_{1}(t) \sqrt{1-R_{1}(t)}|0_{\rm A},1_{\rm B}\rangle\right)\left( \bar{\alpha}_{1}(t)\langle1_{\rm A},0_{\rm B}| +\bar{\beta}_{1}(t) \sqrt{1-R_{1}(t)}\langle0_{\rm A},1_{\rm B}| \right) \\
    & \qquad + |\beta_{1}(t)|^2 R_{1}(t) |0_{\rm A},0_{\rm B}\rangle\langle0_{\rm A},0_{\rm B}|\,,\\
    \rho_{out}^{(2)}(t) 
    & = \left( \alpha_{2}(t)|1_{\rm D},0_{\rm E}\rangle +\beta_{2}(t) \sqrt{1-R_{2}(t)}|0_{\rm D},1_{\rm E}\rangle\right)\left( \bar{\alpha}_{2}(t)\langle1_{\rm D},0_{\rm E}| +\bar{\beta}_{2}(t) \sqrt{1-R_{2}(t)}\langle0_{\rm D},1_{\rm E}| \right) \\
    & \qquad + |\beta_{2}(t)|^2 R_{2}(t) |0_{\rm D},0_{\rm E}\rangle\langle0_{\rm D},0_{\rm E}|\,,\\
    \rho^{(mix)}(t) 
    & =  \langle 0_{\rm C}| U_1(t)\left( |\psi_{in}^{(1)}\rangle\otimes| 0_{\rm C}\rangle\right)  \left( \langle \psi_{in}^{(2)}| \otimes \langle 0_{\rm F}| \right)U_2(t)^\dagger | 0_{\rm F}\rangle \\
    &=\left( \alpha_{1}(t)|1_{\rm A},0_{\rm B}\rangle +\beta_{1}(t) \sqrt{1-R_{1}(t)}|0_{\rm A},1_{\rm B}\rangle\right)\left( \bar{\alpha}_{2}(t)\langle1_{\rm D},0_{\rm E}| +\bar{\beta}_{2}(t) \sqrt{1-R_{2}(t)}\langle0_{\rm D},1_{\rm E}| \right) \,.
\end{split}
\end{equation}
We can see that the overall density matrix is composed of the classical terms $\rho_{out}^{(1/2)}(t)$, which describe the individual PQMs, plus the mixing term $\rho^{(mix)}(t)$ due to the quantum coherence between the two PQMs' evolution. This arises from the coherent splitting before the tunable beam-splitters, which modulate the input fluxes.\\
The average input photon numbers in mode B/E are 
\begin{equation}
    \langle N_{in}^{(1/2)}(t)\rangle \equiv {\rm Tr}\left[ \rho_{in} \mathbf{1}_{\rm A}\otimes\mathbf{1}_{\rm D}\otimes|1\rangle_{\rm B/E}\langle 1|_{\rm B/E}\otimes \mathbf{1}_{\rm E/B}\right] =|\beta_{1/2}(t)|^2 \,,
\end{equation} 
and the corresponding output average photon numbers are
\begin{equation}
    \langle N_{out}^{(1/2)}(t) \rangle \equiv {\rm Tr}\left[ \rho_{out} \mathbf{1}_{\rm A}\otimes\mathbf{1}_{\rm D}\otimes|1\rangle_{\rm B/E}\langle 1|_{\rm B/E}\otimes \mathbf{1}_{\rm E/B}\right] = [1 - R_{1/2}(t)] \langle N_{in}^{(1/2)}(t) \rangle \,.
\end{equation}
The circuit does not allow to measure observables which depend on $\rho^{(mix)}(t)$, since the output channels of the two memristors are not connected.
Thus, the system is characterized by correlations due to the mixing term of the density matrix, and such correlations can be used as an additional resource for the network.

For a weak classical coherent light, the initial state is given by a product of coherent state, described by the product of two displacement operators, each associated with the input state of one PQM. This state and the associated evolution are separable and do not show interference term of the form $\rho^{(mix)}$ at the output of the two parallel PQMs.

\newpage

\printbibliography

\end{document}